\documentclass[twocolumn]{aastex7}

\usepackage{makecell}
\usepackage{graphicx}
\usepackage{amsmath}
\usepackage[encapsulated]{CJK}
\usepackage{ucs}
\usepackage{chngcntr}
\graphicspath{{./}{figs/}}

\newcommand{\prospector}{\texttt{Prospector}}

\newcommand{\msun}{{\rm M}_{\odot}}
\newcommand{\zsolar}{{\rm Z}_{\odot}}
\newcommand{\mstar}{{\rm M}_{\star}}

\newcommand{\ha}{H$\alpha$}

\shorttitle{Population Models as a Solution to Outshining}
\shortauthors{Wang et al}

\begin{document}
\begin{CJK*}{UTF8}{gbsn}

\title{Population Models for Star Formation Timescales in Early Galaxies: The First Step Towards Solving Outshining in Star Formation History Inference}

\correspondingauthor{Bingjie Wang}
\email{bwang@psu.edu}

\author[0000-0001-9269-5046]{Bingjie Wang (王冰洁)}
\affiliation{Department of Astronomy \& Astrophysics, The Pennsylvania State University, University Park, PA 16802, USA}
\affiliation{Institute for Computational \& Data Sciences, The Pennsylvania State University, University Park, PA 16802, USA}
\affiliation{Institute for Gravitation and the Cosmos, The Pennsylvania State University, University Park, PA 16802, USA}
\email{bwang@psu.edu}

\author[0000-0001-6755-1315]{Joel Leja}
\affiliation{Department of Astronomy \& Astrophysics, The Pennsylvania State University, University Park, PA 16802, USA}
\affiliation{Institute for Computational \& Data Sciences, The Pennsylvania State University, University Park, PA 16802, USA}
\affiliation{Institute for Gravitation and the Cosmos, The Pennsylvania State University, University Park, PA 16802, USA}
\email{joel.leja@psu.edu}

\author[0000-0002-7570-0824]{Hakim Atek}
\affiliation{Institut d'Astrophysique de Paris, CNRS, Sorbonne Universit\'e, 98bis Boulevard Arago, 75014, Paris, France}
\email{hakim.atek@iap.fr}

\author[0000-0001-5063-8254]{Rachel Bezanson}
\affiliation{Department of Physics \& Astronomy and PITT PACC, University of Pittsburgh, Pittsburgh, PA 15260, USA}
\email{rachel.bezanson@pitt.edu}

\author[0000-0001-8174-317X]{Emilie Burnham}
\affiliation{Department of Astronomy \& Astrophysics, The Pennsylvania State University, University Park, PA 16802, USA}
\email{efb5552@psu.edu}

\author[0000-0001-8460-1564]{Pratika Dayal}
\affiliation{Kapteyn Astronomical Institute, University of Groningen, P.O. Box 800, 9700 AV Groningen, The Netherlands}
\email{p.dayal@rug.nl}

\author[0000-0002-1109-1919]{Robert Feldmann}
\affiliation{Department of Astrophysics, University of Zurich, Winterthurerstrasse 190, CH-8057 Zurich, Switzerland}
\email{robert.feldmann@uzh.ch}

\author[0000-0002-5612-3427]{Jenny E. Greene}
\affiliation{Department of Astrophysical Sciences, Princeton University, Princeton, NJ 08544, USA}
\email{jgreene@astro.princeton.edu}

\author[0000-0002-9280-7594]{Benjamin D. Johnson}
\affiliation{Center for Astrophysics $\vert$ Harvard \& Smithsonian, Cambridge, MA 02138, USA}
\email{benjamin.johnson@cfa.harvard.edu}

\author[0000-0002-2057-5376]{Ivo Labbe}
\affiliation{Centre for Astrophysics and Supercomputing, Swinburne University of Technology, Melbourne, VIC 3122, Australia}
\email{ilabbe@swin.edu.au}

\author[0000-0003-0695-4414]{Michael V. Maseda}
\affiliation{Department of Astronomy, University of Wisconsin-Madison, Madison, WI 53706, USA}
\email{maseda@astro.wisc.edu}

\author[0000-0003-2804-0648]{Themiya Nanayakkara}
\affiliation{Centre for Astrophysics and Supercomputing, Swinburne University of Technology, Melbourne, VIC 3122, Australia}
\email{wnanayakkara@swin.edu.au}

\author[0000-0002-0108-4176]{Sedona H. Price}
\affiliation{Department of Physics \& Astronomy and PITT PACC, University of Pittsburgh, Pittsburgh, PA 15260, USA}
\email{sedona.price@pitt.edu}

\author[0000-0002-1714-1905]{Katherine A. Suess}
\affiliation{Department for Astrophysical and Planetary Science, University of Colorado, Boulder, CO 80309, USA}
\email{wren.suess@colorado.edu}

\author[0000-0003-1614-196X]{John R. Weaver}
\affiliation{Department of Astronomy, University of Massachusetts, Amherst, MA 01003, USA}
\email{jweaver@astro.umass.edu}

\author[0000-0001-7160-3632]{Katherine E. Whitaker}
\affiliation{Department of Astronomy, University of Massachusetts, Amherst, MA 01003, USA}
\affiliation{Cosmic Dawn Center (DAWN), Niels Bohr Institute, University of Copenhagen, Jagtvej 128, K{\o}benhavn N, DK-2200, Denmark}
\email{kwhitaker@astro.umass.edu}

\begin{abstract}

JWST have revealed temporarily-quenched and ultraviolet-luminous galaxies in the early universe, suggesting enhanced star formation stochasticity. Verifying this hypothesis is critical, yet challenging. Outshining, wherein light from young stars dominates the spectral energy distribution, represents perhaps the greatest challenge in inferring the formation histories of unresolved galaxies. In this paper, we take a simple model of burstiness and show that state-of-the-art inference methods with flexible star formation histories (SFHs) and neutral priors, while recovering average star formation rates (SFRs; $\sim0.1$~dex median offset), fail to recover the complexities of fluctuations on tens of Myr timescales, and typically underestimate masses in bursty systems ($\sim0.15$ dex). Surprisingly, detailed SFH recovery is still sensitive to priors even when data quality is optimal, e.g., including high signal-to-noise ($\rm20~pixel^{-1}$) spectroscopy with wide coverage (rest-frame $0.12-1.06~\mu$m). Crucially, however, refitting the same data with a prior correctly encoding the bursty expectation eliminates these biases: median offsets in mass and SFRs decrease to $\sim 0.04$ dex and $\sim 0.05$ dex, respectively. Under the assumption that current population burstiness predicts past SFH, the solution to outshining in modeling statistical samples is empirically measuring recent galaxy SFHs with population modeling. A prototype is \ha/UV: while helpful, it is insufficient to constrain the expected complex burstiness. To this end, we introduce a more complete, quantitative population-level approach and demonstrate that it promises to recover the typical amplitude, timescale, and slope of the recent SFH to high accuracy. This approach thus has the strong potential to solve outshining using observations from JWST.

\end{abstract}


\section{Introduction}

The growth of galaxies is regulated by physical processes operating across different timescales, ranging from the creation and destruction of giant molecular clouds at below $\sim$~10 Myr to quenching at over $\sim 10^3$ Myr (see \citealt{McKee2007} for a theoretical overview).
Understanding the observed diversity of the galaxy populations---a cumulative effect of these intertwined individual processes---has long been a focus in the field of galaxy formation and evolution.

Theoretical studies based on cosmological simulations offer a promising approach to dissect the formation and evolution of galaxies, but face the challenge of the wide dynamic range required to describe the physics involved. While models agree on a core set of critical processes shaping galaxy properties, their phenomenological implementations (sub-grid physics) differ, leading to a dispersion in model predictions \citep{Somerville2015,Naab2017}. 
Observationally, the imprints of these individual processes are anticipated to shape the star formation histories (SFHs) of galaxies, with each process exerting its influence over distinct characteristic timescales (e.g., \citealt{Iyer2020}). This means that by constraining the variability of SFHs, it would be possible to disentangle the various mechanisms driving or suppressing star formation.

Recent breakthroughs facilitated by the James Webb Space Telescope (JWST) suggest a staggering diversity of inferred SFHs within galaxy populations at early times, reigniting interest in the study of its variability (e.g., \citealt{Looser2023:sfh,Witten2024}). The discovery of mini-quenched galaxies at early epochs \citep{Looser2023:qg,Strait2023} and the over-abundance of luminous high-redshift galaxies have challenged theoretical models (e.g., \citealt{Harikane2023,Mason2023}). 
Large amplitude, short timescale fluctuations in star formation rates (SFRs) have been suggested in $z>1$ studies \citep{vanderWel2011,Atek2011,Maseda2018}. This phenomenon may be prominent in the gas-rich $z>4$ universe, and has been proposed to explain these exotic objects \citep{Pallottini2023,Shen2023,Sun2023,Faisst2024}, in addition to several other theoretical possibilities, including an evolving initial mass function (IMF; \citealt{Cueto2024,vanDokkum2024,Yung2024}), a decrease in dust attenuation with increasing redshift \citep{Mauerhofer2023,Ferrara2024}, a high efficiency of gas conversion into stars \citep{Dekel2023,Renzini2023,Sipple2024}, and black hole contribution to the UV luminosity \citep{Ono2018,Pacucci2022}. This stochasticity is predicted by theoretical models of galaxy formation at high redshift as the feedback timescale approaches or becomes smaller than the dynamical time of the system, disrupting the otherwise simple correlation between smooth gas accretion and SFRs \citep{Stinson2007,Tacchella2016,2017MNRAS.470.1050F,Faucher-Gigueere2018,2023MNRAS.525.5388B,Hopkins2023:disc,Dome2024}. Short-lived and frequent deviations from an ``average" SFR would naturally explain both unexpected low-mass quiescent systems as post-starbursts and overly luminous galaxies as starbursts.

The standard methodology for constraining SFH variability is via \ha-to-UV flux ratios, as the two are expected to trace star formation on different timescales \citep{Weisz2012,Johnson2013,Sparre2017,Caplar2019,Faisst2019,FloresVelazquez2021}. \ha\ emission arises from recombination in nebular clouds, primarily driven by O and B-type stars on timescales of $\lesssim$ 5 or 10~Myr. UV emission is generated by O and B-type stars with lifetimes of $\lesssim 300$~Myr \citep{Kennicutt1998}, although due to outshining of the older stellar light by recent star formation (e.g., \citealt{Papovich2001}), the UV emission in star-forming galaxies typically traces $30-70$~Myr timescales. Pre-JWST studies at $z<3$ have demonstrated the feasibility of this method (e.g., \citealt{Lee2009,Emami2019}), but it faces two main challenges: (i) a purely rising SFH would lead to maximal \ha\ and UV fluxes similar to a bursty SFH \citep{Mehta2023}; (ii) translating the \ha-to-UV flux ratios to timescales is difficult as the timescale probed by the UV is a strong function of the intrinsic SFH, meaning that every object would need to be uniquely and explicitly modeled (e.g., \citealt{Johnson2013,FloresVelazquez2021}).

In principle, one can solve both these problems by performing individual spectral energy distribution (SED) fits instead. Indeed, with advancements in panchromatic SED fitting techniques, numerous studies attempt to estimate SFH variability by fitting photometric observations from deep JWST surveys (e.g., \citealt{Ciesla2023}), in addition to using the \ha-to-UV flux ratios \citep{Asada2024,Endsley2024}. 
However, the challenge here is that the inference of SFHs from individual SED fitting of photometric data, at least in broad-band photometry, cannot distinguish between an ``extremely bursty" and a ``smooth" model. \citet{Wang2023:sys} has demonstrated that both models can produce statistically acceptable solutions, as the Bayes factor shows no preference for either model.

Fundamentally, the challenge is in solving outshining. To infer SFH from integrated light is to infer the past from the present, but with asymmetric information: a recent burst of star formation dramatically outshines the older, fainter stellar populations, even in the near-infrared where old stars are relatively brightest. This prevents an accurate inference of the full SFH, introducing biases in the stellar mass and SFR estimates. \citet{Papovich2001} analyzed the HST photometric data of a sample of Lyman-break galaxies at $2 \lesssim z \lesssim 3.5$, and found that a hypothetical old stellar population could in principle contain up to $\approx 3-8$ times the stellar mass of the young stars that dominate the observed SED; \citet{Wang2023:sys} showed that highly stochastic SFHs constrained only by broadband photometry produce $\sim 0.8$ dex systematics in SFR averaged over a short timescale and $\sim 0.3$ dex systematics in average stellar age.
Recent studies performed on cosmological simulations also hint at systematic biases in inferred stellar masses (e.g., \citealt{Narayanan2024,Cochrane2025}).
Owing to the expected increasing prevalence of burstiness at $z \gtrsim 4$ as outlined previously, and paired with the fact that the SFH is degenerate with most key parameters measured from galaxy SEDs, outshining is perhaps the biggest challenge in interpreting the light from galaxies in the early universe.

In addition, galaxy brightness (and therefore detectability in a flux-limited survey) and burstiness are inherently interconnected, as a recent burst makes galaxies bright. It follows that the duty cycle of the SFH variation would be biased in a flux-limited sample; that is, galaxies without a burst at the low-mass end (over a range of $10-100$ in stellar mass according to a study done on the FIRE simulation; \citealt{Sun2023:seen}) are preferentially missed.
This selection effect has broad implications for statistical studies of the galaxy population such as the mass assembly history and global SFH.

All of the above factors motivate the need for a thorough examination of the extent to which burstiness can be observationally constrained.
In this paper, we take a simple model of burstiness---essentially, repeated up-and-down fluctuations in the SFH---and evaluate the accuracy of recovered SFHs for bursty systems inferred with various methods, including state-of-the-art SED fitting of simulated high signal-to-noise (S/N) spectra, and population-level analyses of spectral features. The latter is motivated by recent developments in making the population distributions of the parameters of interest, such as photometric redshift or stellar population parameters, as the inference objective, bypassing individual fits \citep{Alsing2024,Li2024}. 
All tests have a quantitative target for recovery, enabled by our parameterization of the SFH. This work provides a comprehensive assessment of burstiness constraints, and critically, sets out a potential path forward for solving outshining.

The structure of this paper is as follows. Section~\ref{sec:sim} presents the SFH model and mock observations. 
Section~\ref{sec:ind_fit} focuses on inferring SFH variations from individual galaxies, using common setups in SED fitting. 
Section~\ref{sec:sed_fit_matchsfh} proceeds to show the improvement on these constraints given a correct SFH prior.
Section~\ref{sec:ha_uv} assesses the performance of the widely used population-level burstiness indicator, the \ha-to-UV flux ratio.
Section~\ref{sec:pop} proposes our population-level approach to complement the \ha-to-UV method, and demonstrates its additional constraining power. 
Section~\ref{sec:discussion} ties all the pieces together to address outshining, and also provides additional discussions on the individual SED fits including the prior influences, and the observability and selection effects due to SFH variations.
We conclude in Section~\ref{sec:concl}.

Where applicable, we adopt the best-fit cosmological parameters from the WMAP 9 yr results: $H_{0}=69.32$ ${\rm km \,s^{-1} \,Mpc^{-1}}$, $\Omega_{M}=0.2865$, and $\Omega_{\Lambda}=0.7135$ \citep{Hinshaw2013}, and a \citet{Chabrier2003} IMF over the mass range of $0.08-120~\msun$. Unless otherwise mentioned, we report the median of the posterior, and 1$\sigma$ error bars are the 16th and 84th percentiles.

\section{Simulations\label{sec:sim}}

\subsection{Models of Star Formation History\label{subsec:sfh_model}}

\begin{figure*} 
 \gridline{
 \fig{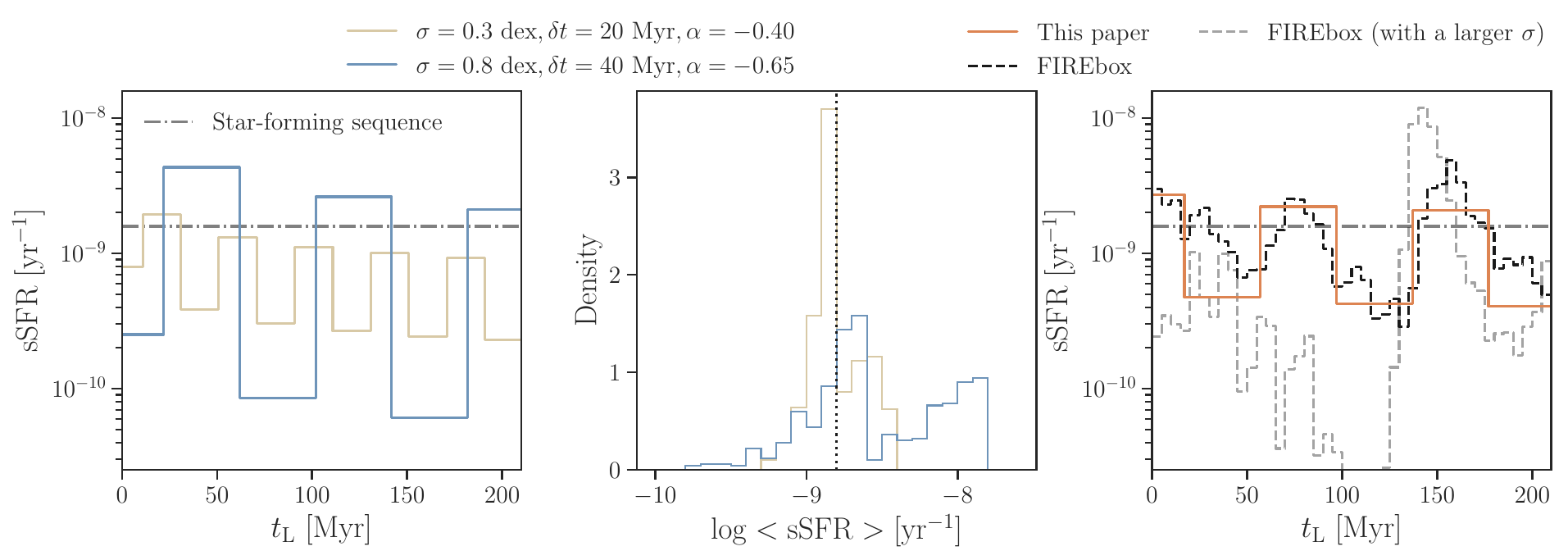}{0.9\textwidth}{}
 }
\caption{Models of star-formation history.
(Left) Two examples of our parameterization of the population-level SFH parameters. One SFH simulated with a fluctuation amplitude $\sigma = 0.3$ dex around the star forming sequence (gray dash-dotted line), a slowly rising slope, $\alpha=-0.40$, and a short duration of the high/low SFR phase, $\delta t = 20$ Myr, is plotted in light brown. One SFH simulated with a greater fluctuation amplitude, duration, and a steeply rising slope is plotted in blue. 
(Middle) The distributions of sSFRs averaged over one duration, for galaxy population observed at different phases, are shown as histograms in the corresponding colors.
(Right) A SFH from the FIREbox simulation \citep{Feldmann2023}, and a parameterized SFH generated for this paper are shown in black and orange, respectively. While our parameterization of the SFH is simple, it captures the key features in many of the FIREbox SFHs. This simplified model provides a clear basis to evaluate the different inference methods. A second SFH, also from FIREbox but with larger fluctuating amplitudes, is over-plotted in gray. The bursts in this example are $\sim 2$~dex peak-to-peak, illustrating that the range of the fluctuating amplitude explored in this work is well motivated.
\label{fig:sfh_model}}
\end{figure*}

\begin{figure*} 
 \gridline{
 \fig{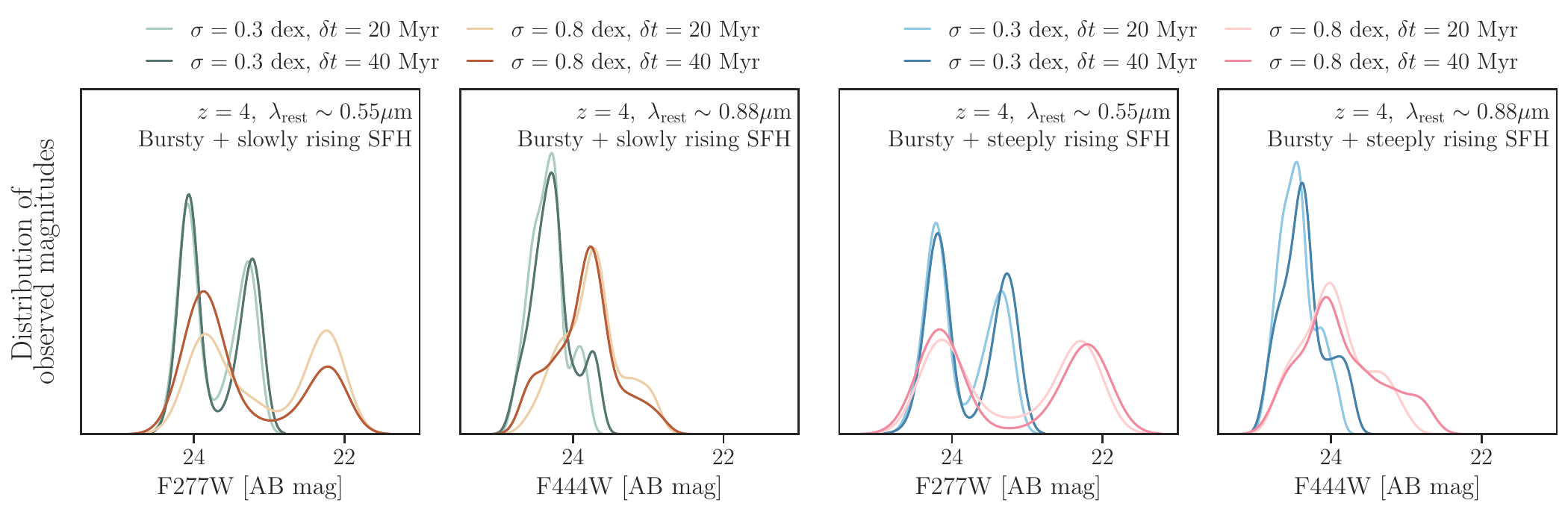}{0.95\textwidth}{}
 }
 \caption{Variations in the F277W and F444W fluxes due to SFH variations alone, with variations due to mass, dust, and redshift removed. For each filter, we include two panels showing the distributions of fluxes from SFHs with a slowly rising slope ($\alpha=-0.40$), and a steeply rising slope ($\alpha=-0.65$). The four curves in each panel reflect the changes in fluxes due to various fluctuating amplitudes and durations. 
For rising SFHs with the same slope, the varying fluctuating amplitudes and durations alone produce over 2 (1.5) magnitudes of variability in the observed rest-frame optical (near-IR) fluxes.
}
\label{fig:f444w}
\end{figure*}

Bursty star formation, defined as repeated up-and-down fluctuations in an otherwise smoothly-varying SFH, is a generic prediction in various simulations of galaxy formation (e.g., \citealt{Stinson2007,Governato2012,Hopkins2014,Faucher-Gigueere2018,Legrand2022}). An example from the FIREbox cosmological volume simulation \citep{Feldmann2023} is shown in Figure~\ref{fig:sfh_model}. Motivated by these SFHs, we develop a simple parameterization, emulating the recurrent ``on'' and ``off'' phases in star formation. It is defined by the following population-level SFH parameters, describing the population distribution from which individual galaxies can be drawn for testing purposes.
\begin{enumerate}
  \item $\sigma$: amplitude of fluctuations in specific star-formation rates (sSFR) around the star-forming sequence, measured in dex.
  \item $\delta t$: duration of the high/low SFR phase, measured in linear time.
  \item $\alpha$: the power-law slope of the SFR during the most recent 500 Myr in lookback time, i.e., $\log({\rm SFR}) \propto \alpha \log(t) $.
  \item $\phi$: phase; $\in$ [0, 2$\pi$], with 0 ($\pi$) referring to the instant when the galaxy is observed at the end of a relatively quiescent (star-forming) period.
\end{enumerate}
The above set of parameters also specifies the total mass formed in the most recent 500 Myr. The choice of 500 Myr is arbitrary, and the results do not depend on this particular choice as long as the time scale is sufficiently long ($\gtrsim 200$~Myr). As we would like to remove the effect on the simulations due to varying galaxy masses, while reaching a target sSFR corresponding to the scatter in the observed star-forming main sequence (e.g., \citealt{Whitaker2012,Speagle2014,Leja2022}), we locate any remaining mass in the last bin.

While the above parameterization is simple, it effectively captures the key features of the oscillating patterns present in many FIREbox SFHs. Certainly, the FIREbox simulation predicts SFHs with more complex forms that cannot be fully described by our model. More generally, different cosmological simulations produce distinct SFHs with power at different timescales \citep{Iyer2020}, largely due to different feedback implementations.
It is worth emphasizing that the purpose of our model is not to replicate SFHs from cosmological simulations. Instead, the simplicity of our parameterization is an intentional choice, providing a clear basis from which the inference of SFH variations from observed SEDs via different methods can be straightforwardly evaluated and compared.
As suggested in Figure~\ref{fig:sfh_model}, this model is sufficient for our testing purposes. We will explore more flexible formalisms in the future.

As a pilot study, we explore a few representative points in the parameter space, bracketing a range guided by observations and simulations. These choices are summarized in Figure~\ref{fig:sfh_model}, and explained in more detail below.
\begin{enumerate}
  \item $\sigma$ = 0.3, 0.8 dex. $\sigma = 0.3$ dex represents the population of ``normal" star-forming galaxies, motivated by the 0.3 dex scatter seen in the star-forming sequence \citep{Whitaker2012,Speagle2014,Leja2022}. $\sigma = 0.8$ dex resembles a burstier case. Note that this parameter represents (the half-width of) the peak-to-peak fluctuations in sSFR(t), i.e., not smoothed over some timescale. Its amplitude can thus be greater than the observed scatter in the star-forming sequence, and we use $0.8$ dex as an illustrative contrasting case to the smaller $\sigma$ value.
  \item $\delta t$ = 20, 40 Myr. \citet{Dome2024} find variations in simulations are typically on the order of 40 Myr, whereas \citet{Maseda2014} posit that the bursts need to be $< 50$ Myr for stability purposes, we thus consider these two illustrative cases.
  \item $\alpha$ = $-0.40$, $-0.65$. As the SFRs of a galaxy have to rise with time in order for it to stay on the star-forming sequence, we consider one slowly and one steeply rising slope. $\alpha=-0.40$ ($\alpha=-0.65$) roughly corresponds to a mass-doubling time of 120 (50) Myr.
  \item $\phi$. We Monte Carlo $\phi$ over the full range of [0, 2$\pi$], meaning that the galaxies are observed at random phases in our simulations.
\end{enumerate}
 Again, we stress that the goal of this work is to perform a strict recovery test; that is, to check whether the individual spectrum, or spectral features contain sufficient information to constrain burstiness. The results will not be dependent on whether the adopted SFH parameterization is a truthful representation of the population-level burstiness of real galaxies, as long as the models capture the spectral characteristics of bursty SFHs.

\subsection{Mock Observations\label{subsec:mock}}

Following the above parameterization, we simulate observations of galaxy populations using \prospector\ \citep{Johnson2021}. We adopt the MIST stellar isochrones \citep{Choi2016,Dotter2016} and MILES stellar library \citep{Sanchez-Blazquez2006} from FSPS \citep{Conroy2010}. Stellar metallicity is drawn randomly in the range of $10^{-1.98}$ to $10^{0.19}~\zsolar$. 
Although the focus is on high redshift, super-solar metallicities are included simply to allow for the mock observations to cover a wide parameter space. This is a conservative approach as the varying metallicities act to increase the scatter in timescale-sensitive features.
Nebular emission is included using a pre-computed \texttt{Cloudy} \citep{Ferland2017} grid \citep{Byler2017}. Gas-phase metallicity is fixed to $10^{-0.5}~\zsolar$. The attenuation of the intergalactic medium (IGM) is assumed to follow \citet{Madau1995}. Dust attenuation is set to zero for simplicity---in practice, for star-forming galaxies with rest-optical spectra, it can be reliably inferred with H$\alpha$/H$\beta$.

\begin{figure*} 
 \gridline{\fig{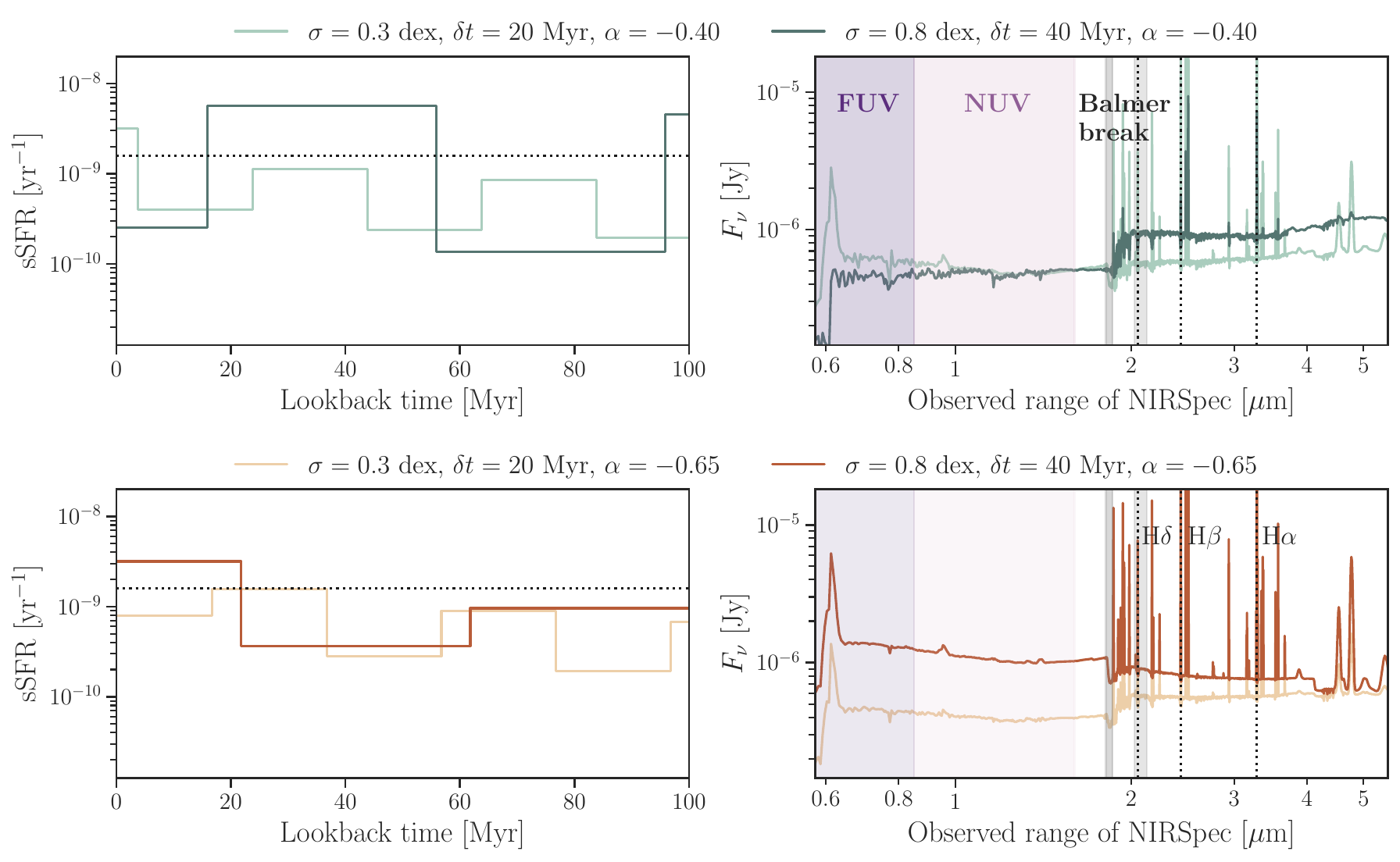}{0.95\textwidth}{}}
\caption{Examples of SFHs and the corresponding model spectra, illustrating the timescale-sensitive spectral features. SFHs simulated with a slowly rising slope of $\alpha=-0.40$, a fluctuation amplitude of $\sigma = 0.3$ dex around the star forming sequence (gray dotted line) and two durations of $\delta t = 20$ Myr and $\delta t=40$ Myr are plotted in light and dark green, respectively in the upper left panel. The resulting model spectra are to the right, with the key observations identified in this paper annotated. The lower panel shows the simulated SFHs and model spectra in a similar format, only that the SFHs assume a steeply rising slope of $\alpha=-0.65$.}
\label{fig:pop}
\end{figure*}

All spectra are shifted to a common redshift of $z=4$. This simplification is not expected to impact our results, since the goal is to test the constraining power of the data. Provided that key timescale-sensitive features, such as the Balmer break and Balmer emission lines, remain present, the exact redshift range is irrelevant for our purpose.
The model spectra, examples of which are shown in Figure~\ref{fig:pop}, are then smoothed with the instrumental line-spread function of the JWST/Prism spectroscopy from JDox\footnote{\url{https://jwst-docs.stsci.edu}}. We include all the JWST NIRCam broad and medium-band photometry, and HST F435W, F606W and F814W filters. This setup is driven by the observing strategy of CANUCS \citep{Willott2022} and UNCOVER \citep{Bezanson2022,Suess2024,Price2024}, which represents the best wavelength coverage for distant galaxies. 
We additionally test the constraining power from the longer wavelengths by including all the MIRI bands, the results of which are presented in the Appendix.
Noise is modeled as Gaussian with S/N $=$ 20, again motivated by the feasibility of obtaining statistical samples with JWST. We simulate 50 galaxies observed at random times for each model to mitigate random measurement uncertainties and sample different phases. 

As an illustration, the distributions of the observed F277W (corresponding to rest-frame optical) and F444W flux (corresponding to rest-frame near-infrared), for a galaxy population at the same redshift and mass, but different SFH variations, are shown in Figure~\ref{fig:f444w}.
The SFH variations alone can produce over one order of magnitude of variability in the observed brightness, driven by the dim objects being in the quiescent phase whereas the bright objects are in the star-bursting phase. 

It follows that star formation variability is critical to the interpretation of flux-limited samples. A lack of understanding in SFH variations translates into a selection bias, where we preferentially observe objects in a starburst. Seen from a different perspective, characterizing star formation variability is also of interest for studies attempting completeness corrections.

\section{Constraining SFH Variations from Individual Fits\label{sec:ind_fit}}

From Figure~\ref{fig:pop}, it appears that different SFHs can have noticeable impacts on the resulting spectra. The most straightforward approach to estimating SFH variations then is via SED fitting of individual galaxies.
We thus begin our analysis by joint photometric and spectral fitting of simulated galaxies, using a setup that follows the most common practices. For ease of comparison, this section contains the methods and results of this individual approach, with further discussions supplied in Section~\ref{subsec:ind}.

We note that the goal is to infer the population-level SFH parameters, as opposed to the usual integrated stellar population properties. For this reason, we conduct this experiment in three stages, described as follows.

\subsection{SED Fitting\label{subsec:sed_fit}}

We perform individual SED fitting using the \prospector\ model \citep{Johnson2021}, representative of the state-of-the-art in Bayesian inference of stellar population parameters (e.g., \citealt{Leja2019:sfh,Tacchella2022:qg}).
We use the same stellar libraries in Section~\ref{subsec:mock}, and the set of free parameters and priors is chosen to follow common practices \citep{Leja2022,Tacchella2022:qg,Wang2024:sps}. 
In brief, the free parameters include the total mass formed, stellar metallicity, 3 parameters controlling the dust based on the two-component model \citep{Charlot2000}, a power-law index for the \citet{Calzetti2000} dust attenuation curve \citep{Noll2009}, and nebular emission described by gas-phase metallicity and an ionization parameter. Nuisance parameters, unconstrained by the available wavelength coverage but included here to be marginalized over, comprise the normalization and dust optical depth of mid-infrared AGN \citep{Leja2018}, and 3 parameters describing dust emission \citep{Draine2007}. 
We follow the standard SED-fitting practice of assuming that all line emission is produced by starlight (without contribution from e.g., shocks or AGN). This means that the strength of the emission lines as probed by our mock spectra directly constrain the short-term SFH.
Altogether, these constitute 13 free parameters.

A 6th-order polynomial is fit to the ratio of the model and observed spectra, so the normalization is set by the photometry, and effectively the shape of the continuum is removed. This ensures that the likelihood is not overly dominated by the continuum shape, allowing the fit to be more sensitive to the spectral features of interest. The polynomial order, however, is low enough to preserve the Balmer break strength, which is also an important age indicator.
Note that the use of the polynomial calibration does not mean disregarding the information encoded in the continuum shape. This is well captured by the excellent photometric data, which includes all the JWST/NIRCam broad- and medium-band filters.

We devote special attention to the assumed SFH forms used in the SED fitting.
As a useful reference point for assessing the recovery of critical stellar population properties such as mass and averaged SFRs, we adopt the widely used \prospector-$\alpha$ model \citep{Leja2017}, in which SFH is described non-parametrically via mass formed in seven logarithmically spaced time bins in lookback time.
A Student's-t prior distribution is placed on the logarithm of SFR ratios between adjacent age bins. The expectation value, $\mu$, of the logarithm SFR ratios is 0. The resulting prior SFH tends toward constant SFR(t), and down-weights dramatic changes between adjacent time bins. The scale, $\sigma$---analogous to the standard deviation for a Gaussian distribution---is 0.3. The allowed prior range is capped at $\pm 10$ for numerical stability. Following \citet{Leja2017}, we refer to this prior as the continuity prior.
We emphasize that this setup is one of the most common ways for using \prospector\ to fit large samples of distant galaxies; therefore our results should be interpreted as applicable in a broader context as well.

The variations in the bursty SFHs at tens of Myr scale necessitate finer time bins. We thus modify the bin scheme in the \prospector-$\alpha$ model.
As a best-case scenario, we tailor our SFH bins to exactly match the periods of quiescence and bursting for each individual galaxy.
In practice, this would be impossible as it requires {\it a priori} knowledge of a galaxy's SFH. While the influence of the choice of the width and location of the age bins has been examined for smoothly varying SFHs in \citet{Leja2019:sfh}, the more complex bursty SFH forms would require a new assessment, especially given that rebinning of a bursty SFH can change the predicted spectrum (see Appendix~\ref{app:rebin}).
Given the possible influence of priors on the inferred SFHs, we consider two Student's-t distributions. First, we use the same continuity prior as proposed in \citet{Leja2017}.
Second, we use the bursty continuity prior from \citet{Tacchella2022:qg}. This differs from the continuity prior on the scale, which is changed from 0.3 to 1, permitting more dramatic changes in SFRs.

In short, we consider three SFH models in this section: \prospector-$\alpha$, which follows the exact definition as in \citet{Leja2017} using 7 time bins; finer time bins matched to the input SFHs, assuming a continuity prior; and the same finer time bins but assuming a bursty continuity prior.
As an additional test, we conduct the same analysis using flexible time bins \citep{Suess2022}. This implementation of non-parametric SFH contains a number of flexible width bins, which is designed to efficiently produce the flexibility required to model post-starburst SFHs. Examples are included in Appendix~\ref{app:grating}.

\begin{figure*}
\gridline{
\fig{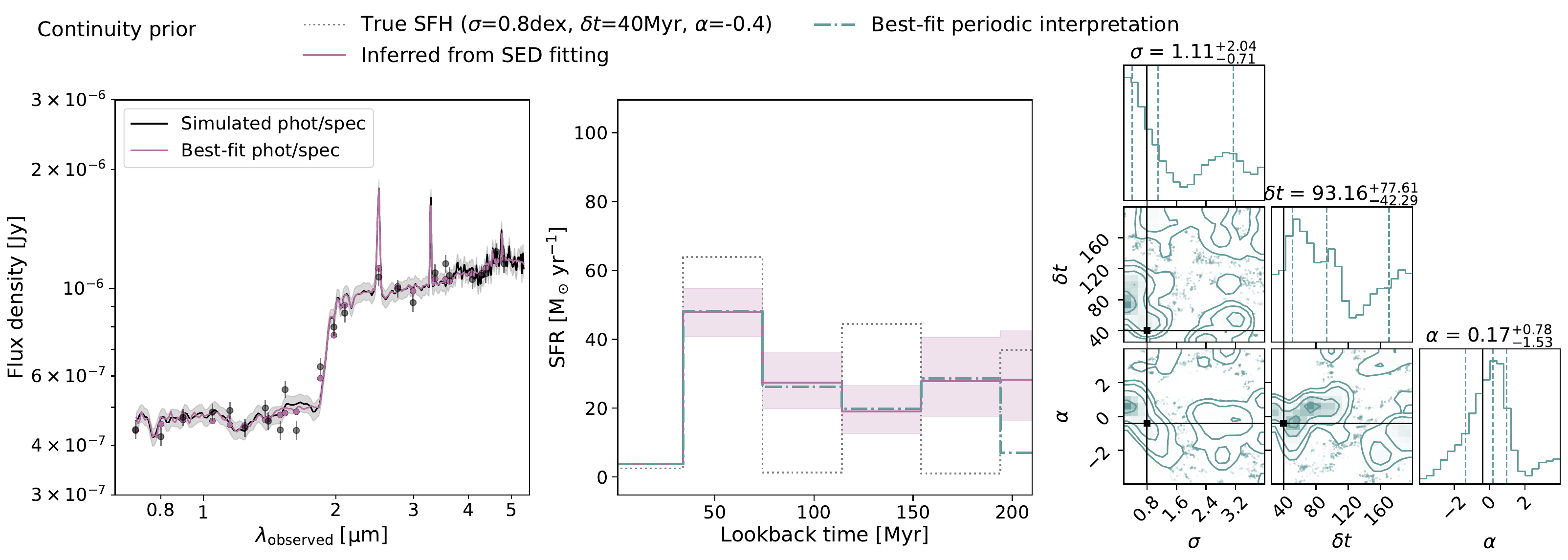}{0.95\textwidth}{(a)}
}
\gridline{
\fig{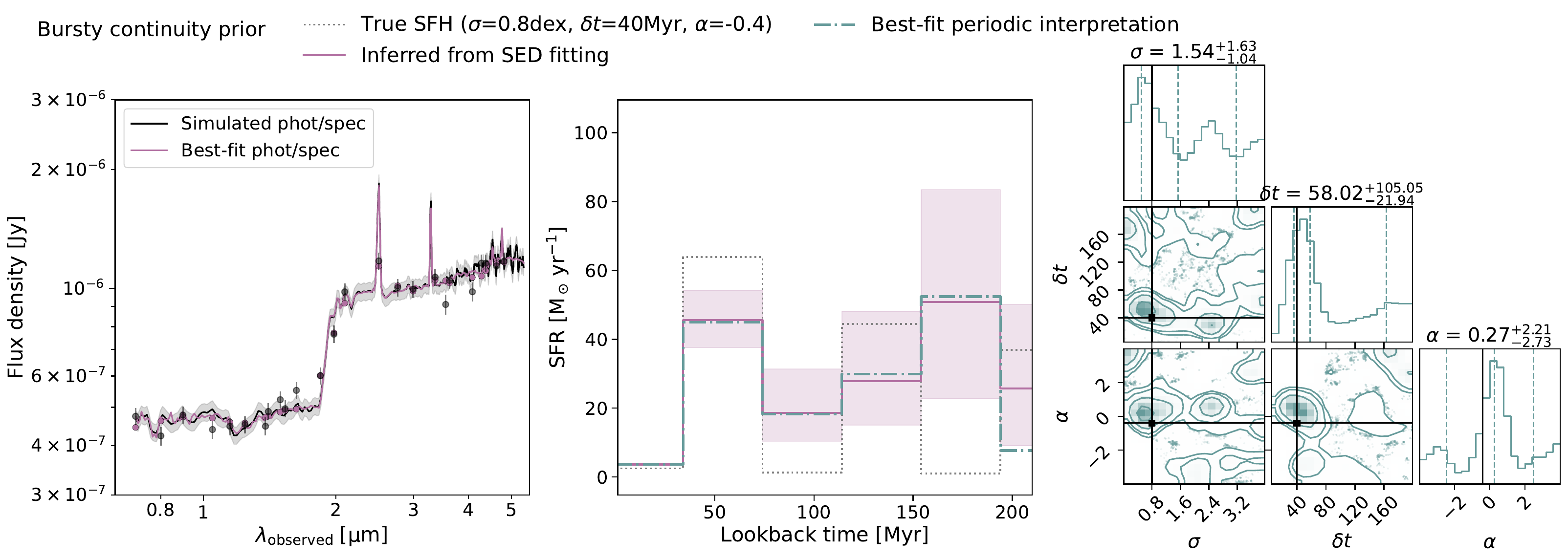}{0.95\textwidth}{(b)}
}
\caption{Individual SED fits. 
(a) An example of a well recovered SFH, assuming a continuity prior in the SED fit. The left panel shows the simulated spectrum in black, and the best-fit model spectrum assuming the continuity SFH prior in purple. 
The middle panel shows the true SFH in black, the non-parametric SFH inferred from SED fit in purple, and its best-fit periodic interpretation in green.
The right panel shows the posterior distributions from the periodic interpretation, with the truths over-plotted in black.
(b) Results for the same galaxy, but assuming the bursty continuity prior are shown in the same manner.
}
\label{fig:ind_sed}
\end{figure*}

\begin{figure*}
\gridline{
\fig{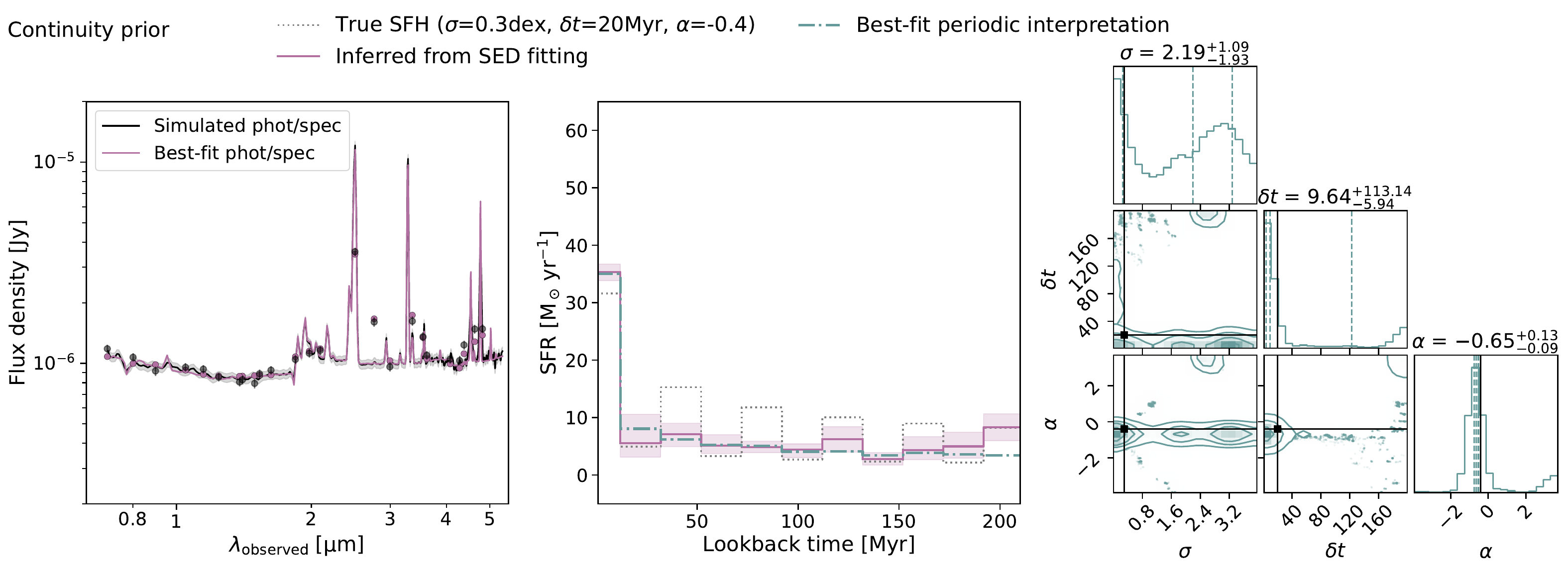}{0.95\textwidth}{(a)}
}
\gridline{
\fig{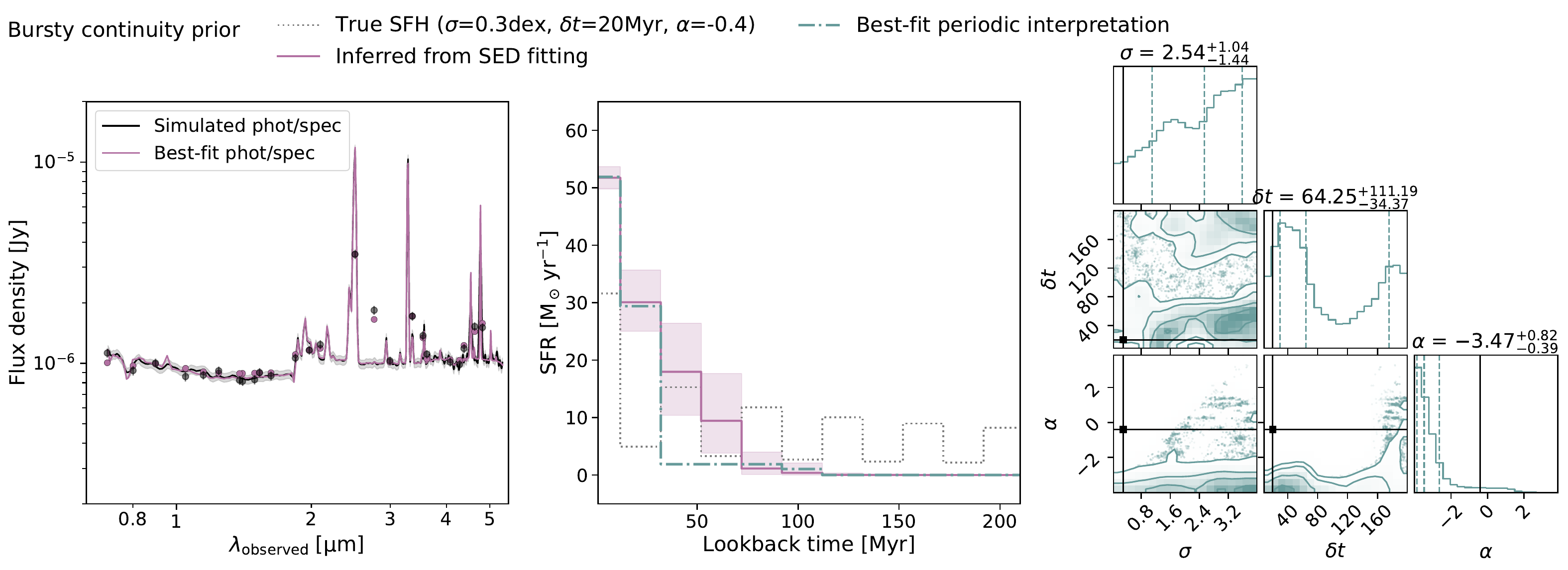}{0.95\textwidth}{(b)}
}
\caption{Individual SED fits. 
An example of a poorly recovered SFH assuming the continuity prior and the bursty continuity prior are shown in the same format as Figure~\ref{fig:ind_sed_bursty}.
The individual fits give noisy information. Even though the spectra are well-fit in all cases, the fidelity of the SFH recovery varies. 
Occasionally, the continuity prior tends to recover the average SFH better than the bursty continuity prior, likely because it prefers less dramatically varying SFHs.
}
\label{fig:ind_sed_bursty}
\end{figure*}

Finally, sampling is performed using the dynamic nested sampler \texttt{dynesty} \citep{Speagle2020}. The simulated and inferred model spectra, as well as SFHs, for two galaxies are included in Figures~\ref{fig:ind_sed}--\ref{fig:ind_sed_bursty} as examples.

\subsection{Periodic Interpretation of Individual SFHs\label{subsec:ind_periodic}}

During SED fitting, we opt not to directly fit for the functional form of the SFH determined by the population-level SFH parameters. This is a more realistic approach, since the SFHs of real galaxies are not expected to follow a simple form, meaning that more complex variations must be allowed. However, the inference objective of interest is the population-level SFH parameters. Therefore, in this second stage, we obtain a periodic interpretation of individual SFHs by fitting the SFHs corresponding to the posterior medians (with $1\sigma$ uncertainty corresponding to the 16 and 84th percentiles) with realizations of the population-level SFH parameters, again using \texttt{dynesty} \citep{Speagle2020}. 
Examples are also shown in Figures~\ref{fig:ind_sed}--\ref{fig:ind_sed_bursty}.

\subsection{Distribution of Population-level SFH parameters from Individual Fits}

The above two steps are repeated for random realizations of the sampled population-level SFH parameters chosen in Section~\ref{subsec:sfh_model}, with 50 simulated galaxies in each SFH family.
In the final step, we examine the distributions of the posterior medians obtained in Section~\ref{subsec:ind_periodic}.

\subsection{Assessing the Accuracy of Individual Fits\label{subsec:ind_res}}

\begin{figure*}
\gridline{
 \fig{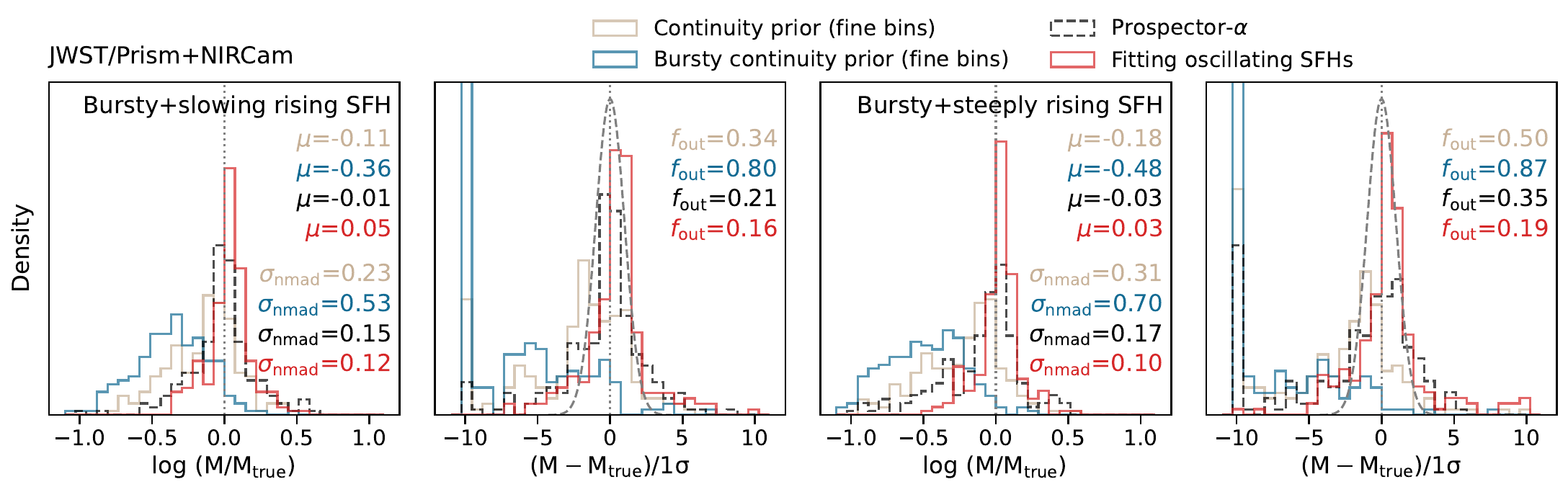}{0.95\textwidth}{(a)}
 }
\gridline{
 \fig{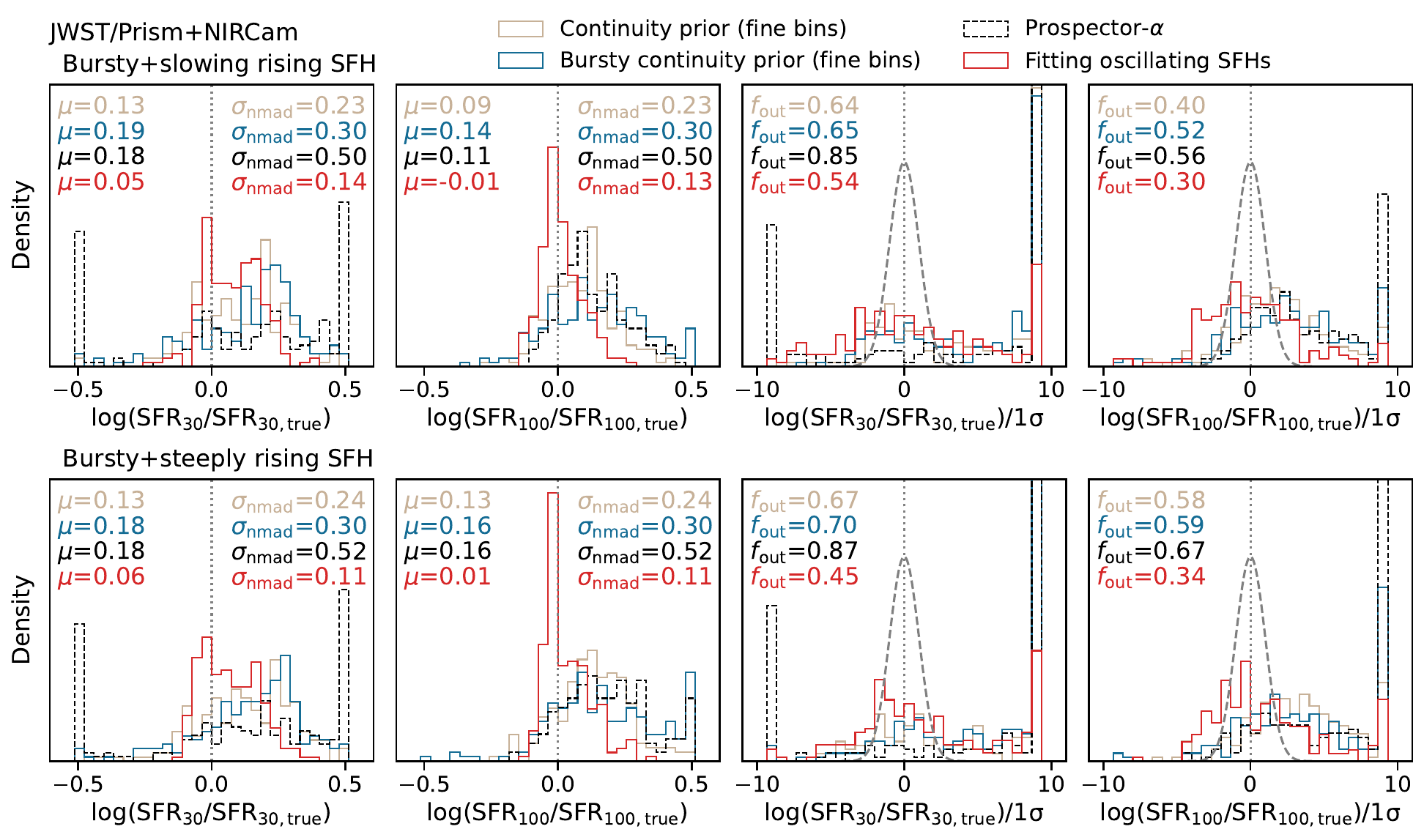}{0.95\textwidth}{(b)}
 }
\caption{Recovery of mass and averaged SFR from SED fitting, under different model assumptions.
(a) The beige and blue histograms show the distributions of difference between the recovered mass and the true mass, assuming the continuity prior and the bursty continuity prior, respectively. The red histograms show the same results using the \prospector-$\alpha$ model \citep{Leja2017}, which describes the SFH in less number of time bins and assumes the continuity prior. This model is included here as a fiducial reference. The black histograms correspond to the results when encoding the correct SFH model in the inference process.
The recovery is quantified by three summary statistics; the median offset, $\mu$, the scatter $\sigma_{\rm NMAD}$, and the outlier fraction, $f_{\rm out}$. These are annotated in each panel in the same colors.
The bursty continuity prior underestimates the mass more often than the continuity prior. This is perhaps due to the bursty continuity prior favoring more dramatic changes in adjacent time bins, leading it to put more mass in a recent burst; it is thus more prone to outshining. Interestingly, the coarse time bins, while insufficient to infer SFH variations on short time scales, lead to less biased mass estimates.
(b) The recoveries of SFRs averaged over the most recent 30 and 100 Myr are depicted in the same format. 
The continuity prior results in a marginally better recovery of the SFR. In all cases, the SFR averaged over the most recent 100 Myr is better calibrated than the SFRs averaged over the most recent 30 Myr.
}
\label{fig:sps_logm}
\end{figure*}

We now evaluate the results of the three previous sections, before concluding with a brief summary. 
Immediately seen from Figure~\ref{fig:ind_sed} is that all the spectra are fit equally well, but the recovery in the critical stellar population properties (mass and averaged SFR) and detailed behaviors in the SFH varies.

\subsubsection{Mass and SFR Recovery}

Histograms showing the recovery of the mass and the SFR averaged over 30, 100 Myr timescale and the corresponding uncertainty calibrations are in Figure~\ref{fig:sps_logm}. The median offsets in the inferred mass, assuming the continuity prior, for the slowly and steeply rising SFHs are -0.11 and -0.18 dex, respectively. These values for the cases assuming the bursty continuity prior, for the slowly and steeply rising SFHs, are -0.36 and -0.48 dex, respectively.
Such underestimations in mass are clear demonstrations of outshining, where the older stellar populations are too faint to be inferred correctly from the integrated light. Using the bursty continuity prior systematically underestimates the mass slightly more, since this prior prefers to locate most of the star formation activity in a recent burst.
Notably, the \prospector-$\alpha$ model, which uses fewer time bins to describe the SFH, is much less prone to underestimate the masses. The mean bias in the inferred masses for slowing (steeply) rising SFHs, is $-0.01$~dex ($-0.03$~dex).

The scatter in residuals is quantified using the normalized median absolute deviation (NMAD), given its advantage of being less sensitive to outliers than root mean square. It is defined as
$\sigma_{\rm NMAD} = 1.48 \times {\rm median} | x_{\rm obs} - x_{\rm true} |$.
The \prospector-$\alpha$ model results in the least scatter in the inferred stellar masses, with $\sigma_{\rm NMAD}\approx 0.16$ dex. With finer time bins, the scatter for the cases assuming the continuity prior and the bursty continuity prior $\approx 0.27$ and 0.62 dex, respectively.

We quantify the uncertainty calibration via an outlier fraction, $f_{\rm out}$, defined as the number of objects where $| x_{\rm obs} - x_{\rm true} | > 3 \sigma$, where $3 \sigma$ refers to the 0.1 and 99.9\% credible interval value distance from the posterior median. In the case of slowly rising SFHs, the outlier fractions for the inferred stellar masses for the \prospector-$\alpha$ model, finer time bins with the continuity prior, and finer time bins with the bursty continuity prior are 0.21, 0.34 and 0.80, respectively. In the case of steeply rising SFHs, these values increase to 0.35, 0.50, and 0.87, respectively.
Interestingly, the standard \prospector-$\alpha$ model \citep{Leja2017} performs the best in all metrics for mass recovery. Consistent with the findings in \citet{Leja2019:sfh}, this non-parametric description with sufficiently coarse time resolutions is able to recover mass across a range of SFH forms.

However, the robust performance in mass recovery with the coarse time bins is not retained in the inference of SFRs. As shown in Figure~\ref{fig:sps_logm}-b, the median offset, scatter, and outlier fraction in SFR averaged over the most recent 30 (100) Myr are 0.2 (0.1) dex, 0.5 (0.5) dex, and 0.9 (0.6), respectively.
Finer age bins with the continuity prior perform marginally better, as it tends to trace the average SFH more accurately. The median offset, scatter, and outlier fraction in SFR averaged over the most recent 30 (100) Myr are 0.1 (0.1) dex, 0.2 (0.2) dex, and 0.6 (0.5), respectively.
Finer age bins with the bursty continuity prior result in similar median offset and scatter as the standard \prospector-$\alpha$ model, except that the outlier fraction is slightly lower (0.7 for SFR averaged over the most recent 30 Myr, and 0.5 for SFR averaged over the most recent 100 Myr).
In all cases, SFRs averaged over a longer time scale (100 Myr) are recovered more accurately than those averaged over a short time scale (30 Myr), at $<0.1$~dex level in median offsets.
This finding is consistent with \citet{Wang2023:sys}, where the systematic uncertainty in the inferred SFRs from broad-band photometry, driven by the modeling choices of a smooth and an extremely bursty SFH prior, is found to be substantially worse for SFRs averaged over short time scales.

Overall, under the finer time bin scheme, the masses and averaged SFRs are reasonably recovered, with $<0.2$ dex offsets. The concerning aspect is that the uncertainties tend to be significantly underestimated, as seen from the large outlier fractions. This is likely due to a complex likelihood surface that is populated with local maxima, creating a unusually challenging situation for the sampler. We examine this point further in Appendix~\ref{app:sampler}.

\subsubsection{Inferring Short-Term Variations in the SFH}

The individual non-parametric SFH can be a poor description to the input SFH.
Earlier studies have shown the influence of priors on the inferred SFHs using photometric data \citep{Leja2019:sfh,Tacchella2022:qg,Wang2023:sys}, but here we demonstrate that their impact can still be significant in the presence of high S/N spectra. We also find that the continuity prior occasionally recovers the SFH averaged over a full period better than the bursty continuity prior, as it prefers more smoothly varying SFHs.

The above challenge translates to a wide dispersion in the quality of the periodic interpretation of individual SFHs (Figures~\ref{fig:ind_sed}--\ref{fig:ind_sed_bursty}). Even in the case of reasonably traced SFH variations from the individual non-parametric SFH fit, it is challenging to estimate the SFH parameters, as illustrated by the wide, multi-modal posterior distributions.
In the case of a poor non-parametric fit, the truths can fall outside of the posterior distributions. The amplitude and the slope are often degenerate, further complicating the interpretation.

\begin{figure*}
\gridline{
  \fig{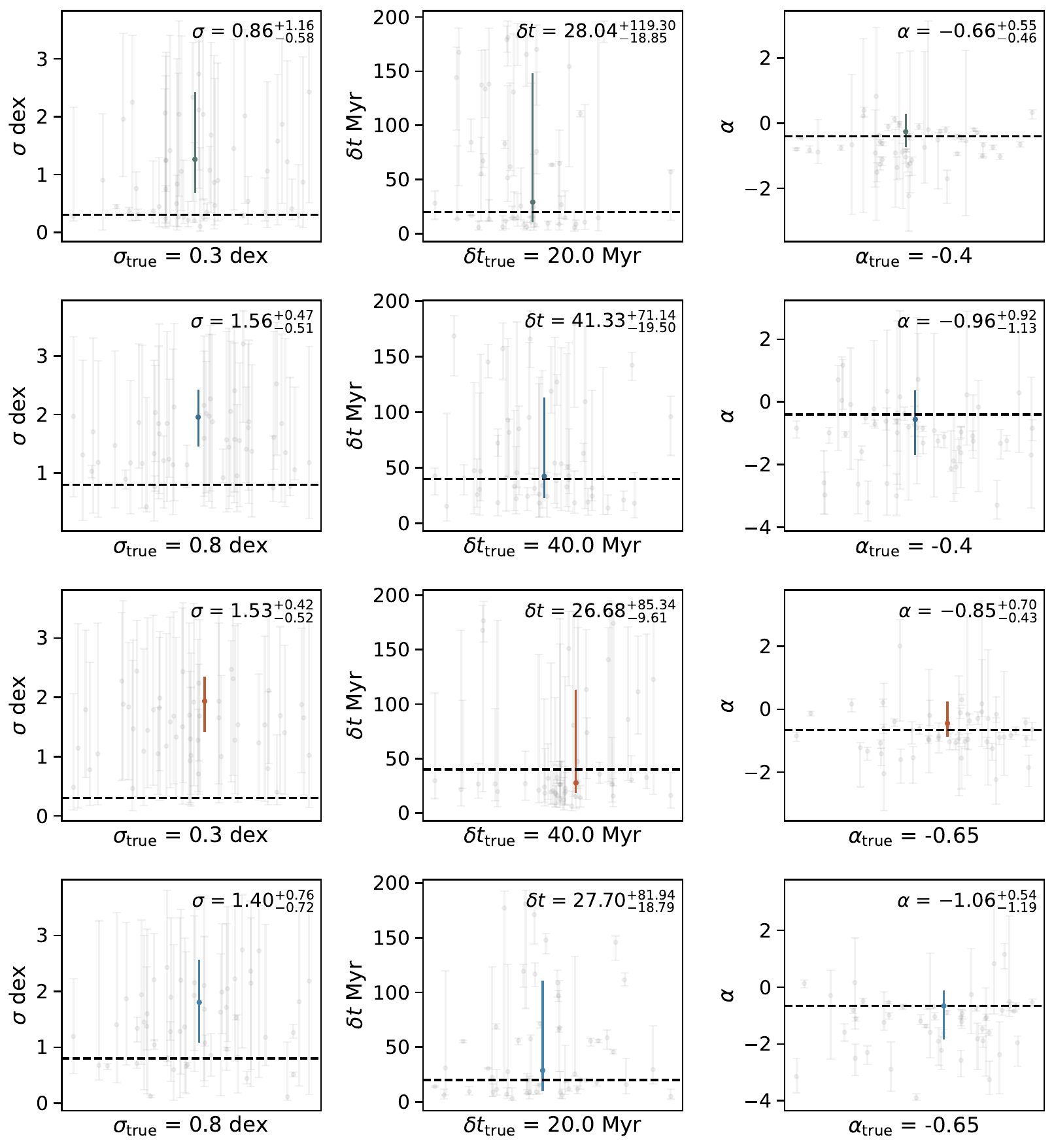}{0.8\textwidth}{}
}
\caption{The accuracy and precision of the inferred population-level SFH parameters from individual fits vary wildly. Each row corresponds to a SFH family. Each panel shows the posterior moments (median / 16th / 84th quantiles) of the population-level SFH parameters, amplitude $\sigma$, duration $\delta t$, and slope $\alpha$, inferred assuming the continuity prior, in gray. The median of the distribution of the posterior medians in gray are shown as a colored data point, with the error bars indicate the 16th and 84th quantiles. The actual numerical values are annotated in the upper right corners.
Truths are indicated as black horizontal lines to guide the eye.
The spread in the posterior medians are often larger than the spacing of the model grid, demonstrating that it is not possible to determine which population model any individual galaxy was drawn from. Notably, some gray points are far from the truth and have small errorbars. This means that if we were to infer the burstiness from these individual fits, we would arrive at the wrong, but confident conclusion that there is a great diversity in the level of burstiness.
}
\label{fig:ind_res_std}
\end{figure*}

\begin{figure*}
\gridline{
  \fig{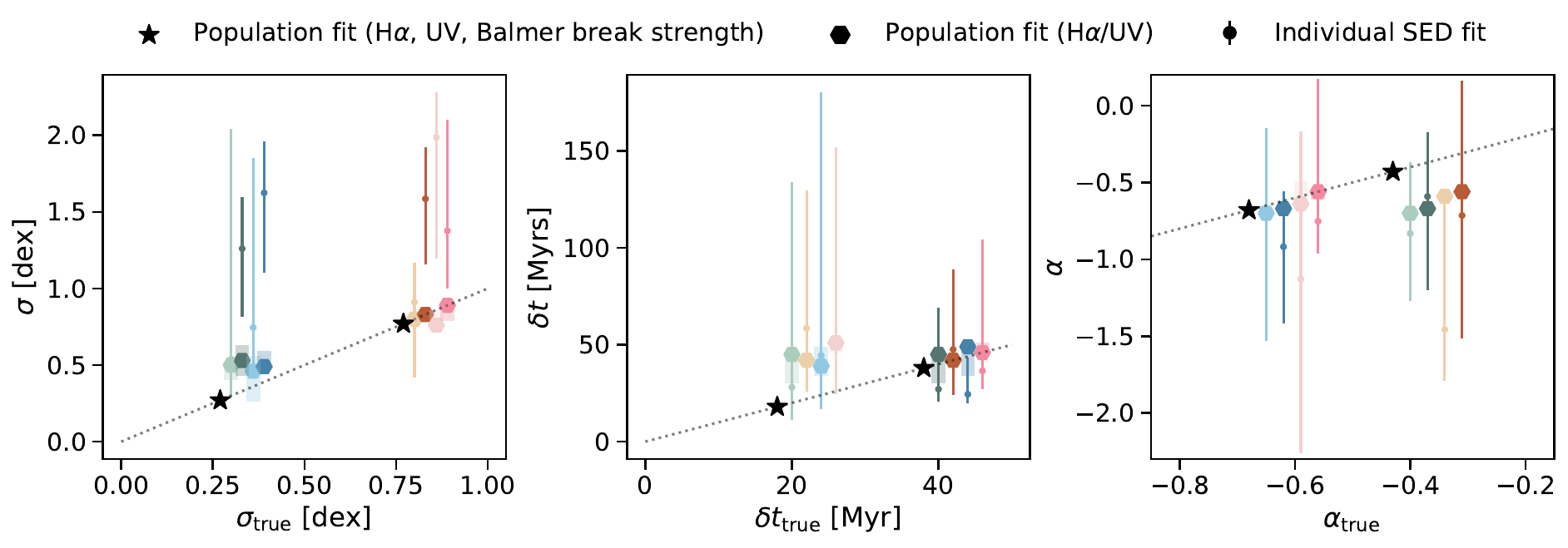}{0.95\textwidth}{}
}
\caption{Contrasting the recoveries of the population-level SFH parameters via individual SED fit (Section~\ref{sec:ind_fit}), \ha/UV (Section~\ref{sec:ha_uv}), and population fit (Section~\ref{sec:pop}).
The colored data points indicate 50th quantiles of the distributions of posterior medians of the population-level SFH parameters, amplitude $\sigma$, duration $\delta t$, and slope $\alpha$, inferred from individual fits, assuming the continuity prior, and the error bars indicate the 16 and 84th quantiles.
The colors correspond to different SFH models.
The $1\sigma$ spread in the posterior medians are often larger than the spacing of the model grid, demonstrating that it is not possible to determine which population model any individual galaxy was drawn from.
The 100\% success rate of identifying the correct model population-level SFH parameters via the population-level approach is illustrated by the stars.
Note here that the error bars mean the spread in posterior medians. The individual fits often have uncertainties far smaller than their biases.
}
\label{fig:comp}
\end{figure*}

Third, the distributions of posterior moments sufficiently illustrate the lack of constraint in the inferred population-level SFH parameters.
The results based on SED modeling assuming the continuity prior are shown in Figure~\ref{fig:ind_res_std}.
The distributions based on the bursty continuity prior exhibit similar trends, albeit with larger deviations from the true values.
A simplified version of Figure~\ref{fig:ind_res_std} is included as Figure~\ref{fig:comp}, which summarizes the results on a scatter plot with error bars indicating the 16 and 84th percentiles of the distributions of the posterior medians.
The accuracy and precision of the inferred population-level SFH parameters from individual fits vary wildly. In most cases, the uncertainties are so large that render the constraints meaningless; i.e., they span the allowed prior range. In some cases of the duration and slope parameters, the answers are wrong with high confidence. The latter is particularly concerning: if we were infer the burstiness from these individual fits, we would arrive at the wrong, but confident conclusion that there is a great diversity in the level of burstiness.
Looking only at the distributions of posterior medians, which are annotated in the right corners of Figure~\ref{fig:ind_res_std}, the spread in the posterior medians are often larger than the spacing of the model grid. For instance, the $1\sigma$ scatters in the inferred amplitudes and durations can be up to 1 dex. This means that our sampled population-level SFH parameters---the periodic bursts of 20 and 40 Myr, the fluctuation amplitudes of 0.3 and 0.8 dex, and doubling times of 50 and 120 Myr---cannot be distinguished.

Simply put, the substantial spread in the inferred population-level SFH parameters limit our ability for an accurate accounting of the population-level burstiness, even though all the spectra are well fit. We therefore conclude that the information content in the spectra is insufficient for inferring the population-level SFH parameters via the state-of-the-art inference methods with flexible SFHs and neutral priors.

\begin{figure*}
 \gridline{
 \fig{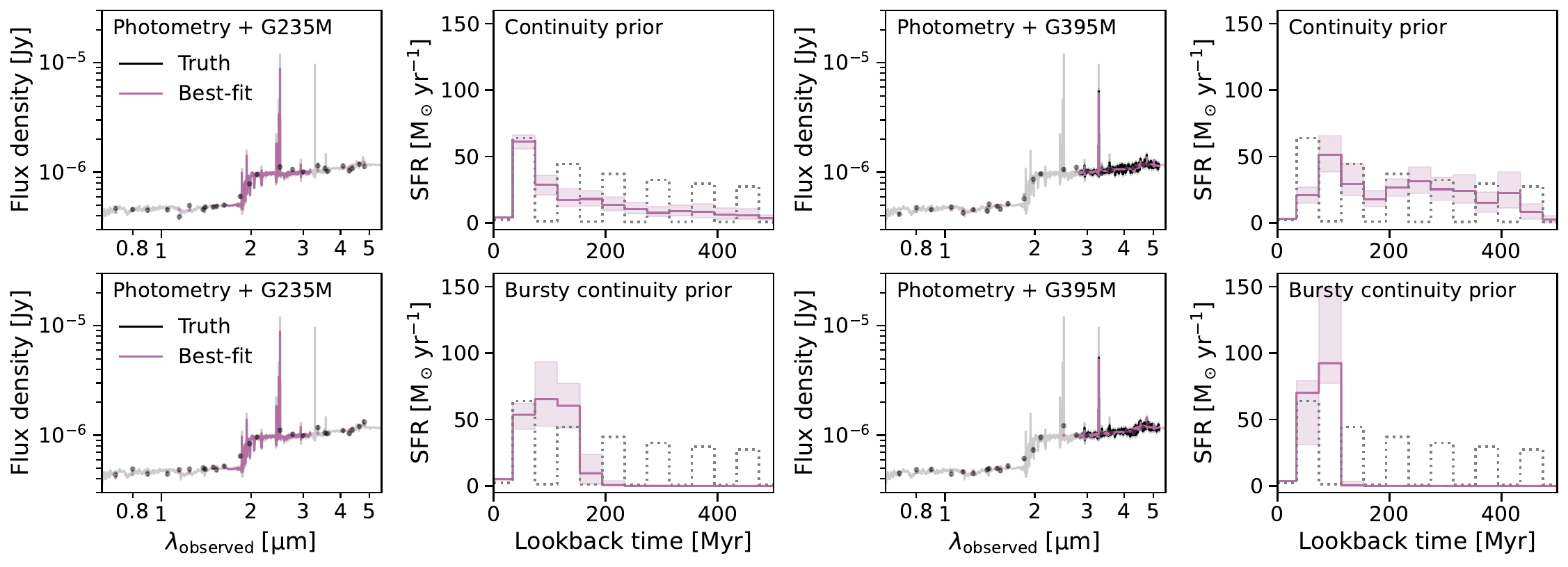}{0.95\textwidth}{}
 }
\caption{Inferring SFHs with medium-resolution spectra. The results are presented in the same format as Figure~\ref{fig:ind_sed}.
It remains difficult to recover the SFH with medium-resolution spectra; despite the excellent fit to the data and systematic-free modeling (impossible with real data!), the input SFH is not recovered.
This reinforces our finding that detailed SFH and therefore its variability often pushes beyond the limit of the state-of-the-art stellar population synthesis modeling.
}
\label{fig:grating}
\end{figure*}

\subsection{Comparing the Accuracy of Individual Fits across Different Data Resolutions\label{subsec:ind_res_other}}

Having focused on recovering the oscillating SFH models using JWST broad and medium-band photometry + Prism spectroscopy, it is also instructive to know whether this is a fundamental problem driven by lack of observational information, or whether higher resolution spectra would provide more constraining power. Therefore, we devote this section to a comparison of information recovery for the individual fits using different and more detailed observations.

We adopt the same general \prospector\ settings as outlined in Section~\ref{subsec:sed_fit}. Again, we stress that our setup is representative of the most common practices in performing SED fits. Two additional observing strategies are considered: photometry + medium-resolution G235M spectroscopy, and photometry + medium-resolution G395M spectroscopy. At our redshift range of interest ($z \sim 4 - 7$), the G235M grating covers a critical age indicator---the Balmer break---and also H$\beta$, whereas the G395M grating covers H$\alpha$ and H$\beta$. 

The results are summarized in Figure~\ref{fig:grating}. In this example, the observable signatures---including the Balmer break, stellar absorption features, and emission lines---are all encapsulated by the G235M grating, and they are distinct enough for the models to correctly infer the recent truncation in star formation. However, despite the excellent fit to the data, the details of the recent SFH are not recovered. The same effects on the inferred SFHs due to the two priors are also seen here; that is, the continuity prior tends to recover the average SFH, whereas the bursty continuity prior tends to locate most of the star formation activity in a recent burst. As for the photometry + G395M grating combination, the medium bands capture the strength of the Balmer break, but the spectral coverage only includes the \ha\ emission. The recent SFH inferred with the continuity prior is less accurate possibly due to the lack of constraints from the rich absorption features around the Balmer break.

Additional comparisons using a rejuvenating SFH, fit with the flexible time bins, produce similar results. Those are presented in Appendix~\ref{app:grating}.

The above results all support the conclusions in Section~\ref{subsec:ind_res} that the information content contained in the individual spectra is, even in the most favorable systematics-free situation, too limited to correctly infer the complex behaviors in the SFH.
However, it remains to be demonstrated how much the individual fits can be improved, given a correct model of burstiness. We do so in the next section.

\section{Individual SED fits, revised with a correct model of burstiness\label{sec:sed_fit_matchsfh}}

\begin{figure*}
 \gridline{
 \fig{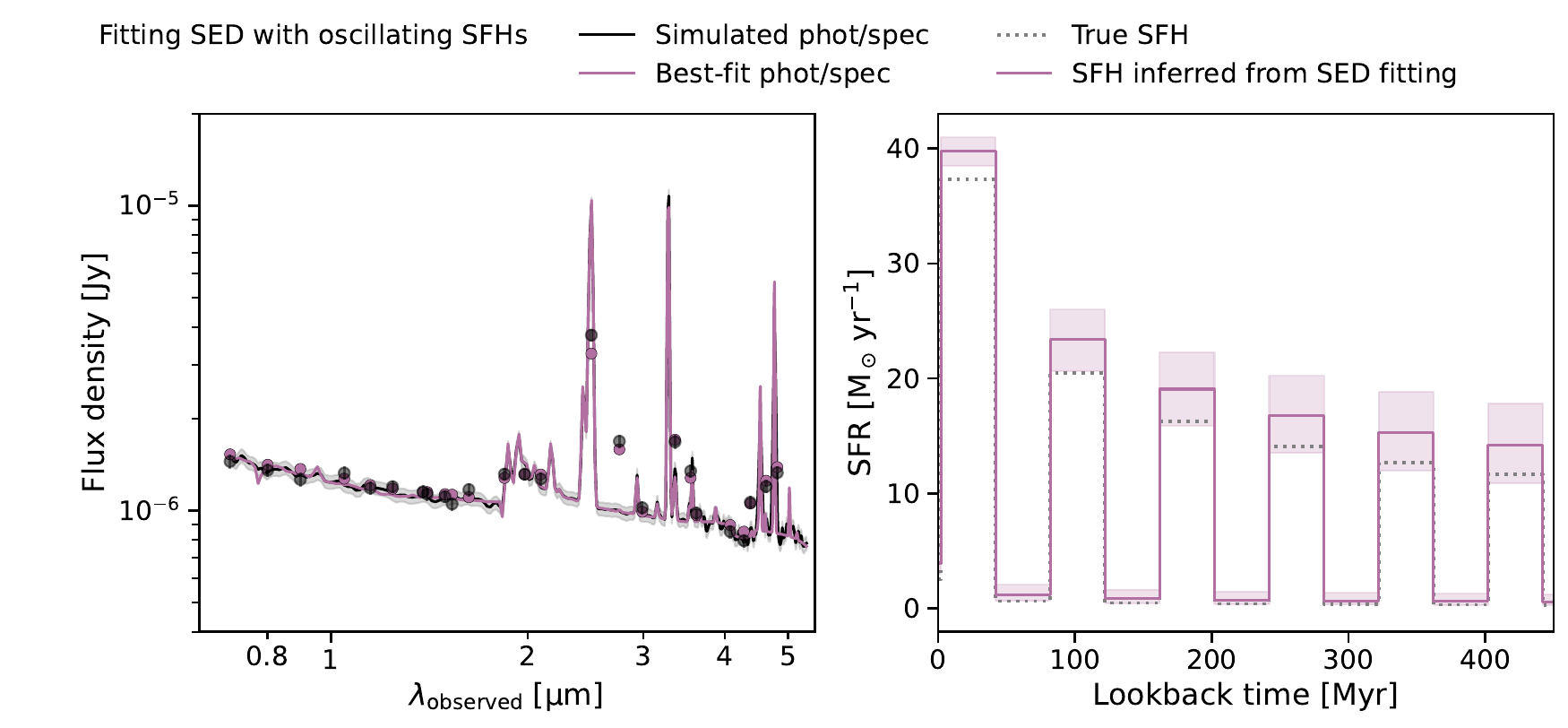}{0.95\textwidth}{(a)}
 }
 \gridline{
 \fig{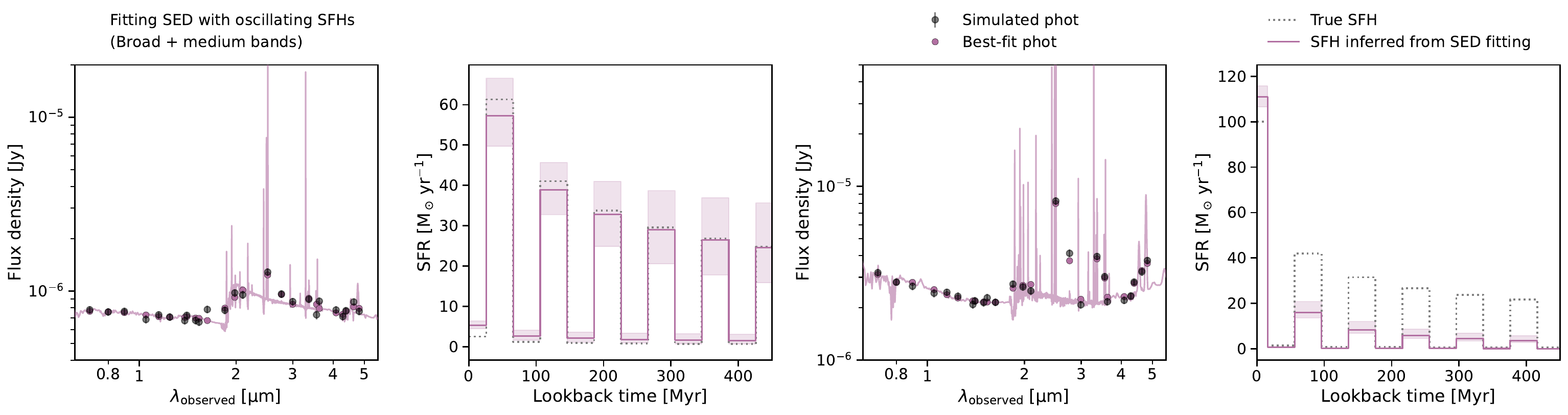}{0.95\textwidth}{(b)}
 }
\caption{With the appropriate model for burstiness, inferring the SFH via SED fitting is indeed feasible. The results here are shown in the same format as Figure~\ref{fig:ind_sed}. (a) Fitting the photometry and Prism spectrum simultaneously with the correct burstiness model recovers the input SFH. Note that we assume no dust attenuation here for simplicity. (b) Under the same settings, fitting only the broad- and medium-band photometric data can produce different performances. Generally, for cases where clear age indicators (e.g., the Balmer break in the left example) are captured by the medium bands, the input SFH can be reasonably recovered without spectroscopy. The x-axis range is truncated to emphasize the recent SFH.\label{fig:sed_matchsfh}}
\end{figure*}

Thus far, we have adopted the common setup (the continuity and bursty continuity priors) used in SED fits, but these state-of-the-art priors do not specifically weight towards the oscillating behavior of the simple SFHs; i.e., the model allows for but does not prefer the oscillating SFHs used as inputs. 
Having demonstrated that even high-quality spectra with a neutral prior cannot recover the detailed SFH, it is natural to ask whether individual SFHs can be accurately inferred with a prior tuned to the underlying population SFH. Such priors may be needed to solve the outshining problem, and can be empirically measured: \citet{Wang2023:pbeta} demonstrated the utility of physically motivated, informative priors in breaking degeneracies among stellar population parameters. More recently, \citet{Gao2024} recovered cosmic SFR density and stellar mass density by constructing a prior based on a correlation between photometric colors and SFHs from the IllustrisTNG simulation \citep{Weinberger2017,Pillepich2018}, whereas \citet{Wan2024} showed the potential for a hierarchical modeling approach with a flexible stochastic SFH prior that aims to describe a wide range of SFH behaviors.

In this section, we test the hypothesis that, if the short-term SFH variability is known {\it a priori}, the SFH of individual galaxies can indeed be recovered via SED fitting. To this end, we fit mock observations using an oscillating SFH model as parameterized in Section~\ref{subsec:sfh_model}. While this choice is intentionally avoided in Section~\ref{subsec:sed_fit}---given that real galaxies are not expected to strictly follow a simple SFH form---it serves as a clear test here to address the question: if the model is informed that the SFH follows an oscillating pattern, can we recover the SFH from fitting individual spectra? Here, we use uniform priors on all four population-level SFH parameters: $\phi \in [0, 2\pi]$, $\sigma \in [0, 1.0]$ dex, $\alpha \in [-1, 0]$, and $\delta t \in [10, 70]$ Myr.

The summary statistics quantifying the stellar population parameter recovery are included in Figure~\ref{fig:sps_logm}. With the correct model for population-level burstiness, the distribution of the residuals in the inferred mass deviates from 0 by $<0.05$ dex, the scatter is $0.1$ dex, and the distribution of uncertainty-normalized residuals becomes close to a unit Gaussian with an outlier fraction of $0.2$.

Furthermore, the intricate variations in the SFH can also be recovered.
An example is shown in Figure~\ref{fig:sed_matchsfh}, where the inferred SFH is consistent with the true SFH. We note that the good recovery is contingent upon an accurate measurement of dust attenuation. This is because dust attenuation is needed to decode the Balmer emission lines, tracing SFR in the most recent $\sim 10$ Myr, and UV fluxes tracing SFR in the most recent $\sim 30 - 70$ Myr scale. The degeneracy between dust attenuation and intrinsic emission-line luminosity, given an observed flux, can cause a factor of $>2$ (up to orders of magnitudes) deviation from the true recent SFRs. Uncertainty in the deblending of \ha\ and [N\,\textsc{ii}] introduces a second order effect where the recent SFH can be incorrectly inferred at a level of $\sim 20$\%, since [N\,\textsc{ii}] typically is $\sim 10$\% of \ha\ flux.

Figure~\ref{fig:sed_matchsfh}-b additionally includes two examples of inferring the SFHs with photometry alone. Generally, for cases where clear age indicators (e.g., a Balmer break) are captured by the medium bands, the input SFH can be recovered without spectroscopy. However, as alluded to earlier, in some cases accurate and precise measurements of the Balmer emission lines are necessary for recovering the SFH.

The above findings suggest that given the correct SFH model, it is indeed possible to infer the SFH by performing SED fitting. Put in another way, outshining can be solved if the right SFH model is known. A key assumption here is that the current burstiness observable in the population is consistent with the past SFH behavior. For cases where star formation is primarily self-regulated and stationary---rather than driven by external events like mergers, or evolving sharply with time due to e.g., gas availability---this assumption is likely reasonable. We revisit this point in Section~\ref{sec:discussion}.

To summarize, the key to address outshining is to empirically measure the population distribution of bursty SFHs. Possible ways forward to achieve this is the focus of the second half of the paper.

\section{\ha-to-UV Flux Ratio\label{sec:ha_uv}}

\begin{figure*}
 \gridline{
 \fig{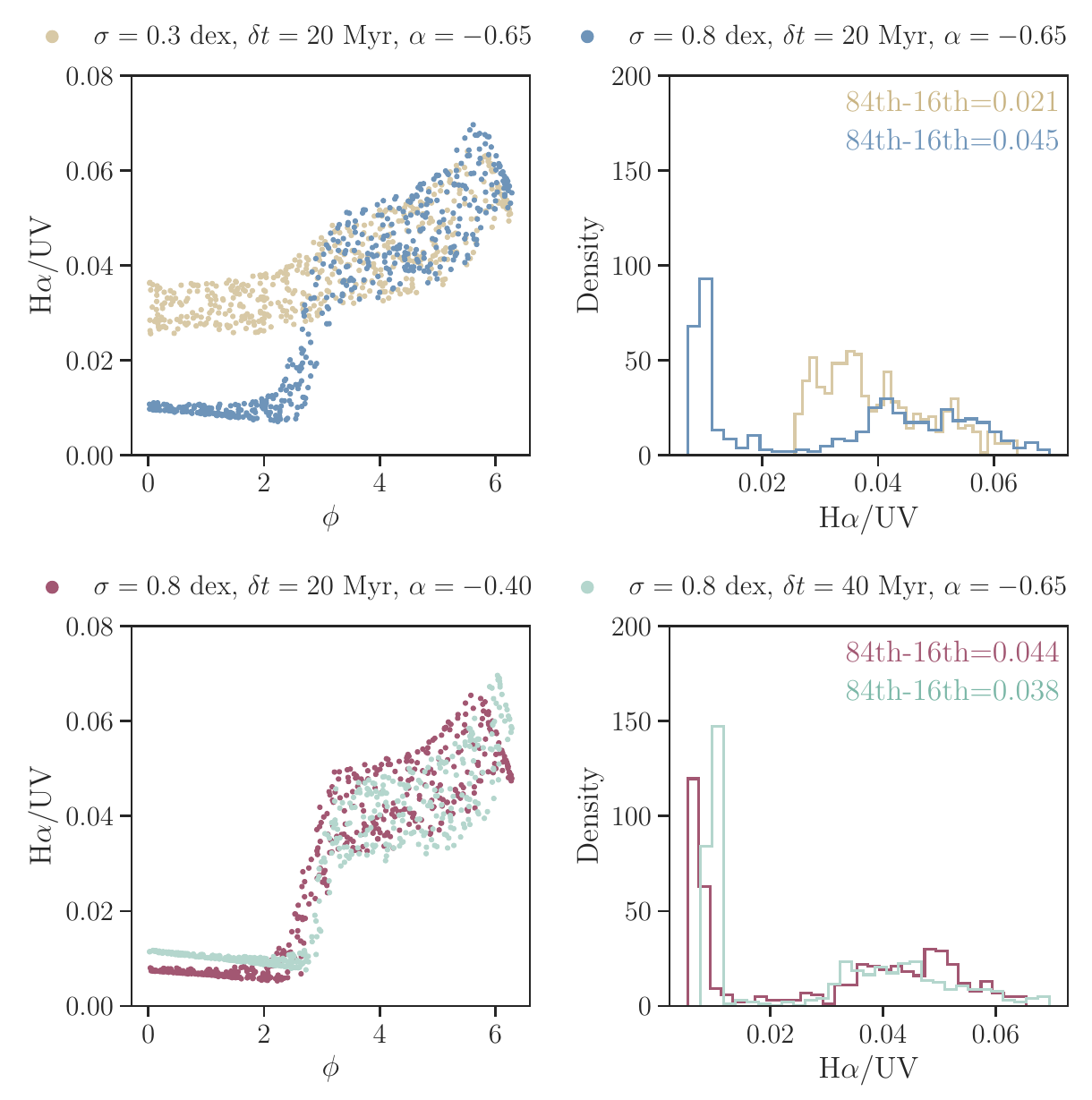}{0.95\textwidth}{}
 }
\caption{\ha-to-UV flux ratios are sensitive to some, but not all population-level SFH parameters. (Upper) The flux ratios are plotted as functions of observed phases in the first panel, whereas the same flux ratios are plotted as histograms to illustrate their distributions, assuming all phases are observed, in the second panel. The two SFH families considered have different fluctuation amplitudes, and the \ha-to-UV flux ratios show different distributions, meaning that they are sensitive to the fluctuation amplitudes. (Lower) Same as the upper panel, but the two SFH families considered here share the same fluctuation amplitude. Although one SFH family has a slowly rising slope and a short duration (purple), and the other SFH family has a steeply rising slope and a longer duration (cyan), the distributions of the flux ratios exhibit marginal difference. 
\label{fig:hauv}}
\end{figure*}

The most well-studied measurement of burstiness is likely the scatter in \ha-to-UV flux ratios. There is rich literature on this variability indicator \citep{Glazebrook1999,Lee2009,Weisz2012,Guo2016,Emami2019,Nanayakkara2020}, and it can be considered as a prototype of a population-level fit. We therefore start the presentation of population-level constraints of burstiness with \ha-to-UV flux ratios, and test its constraining power using our mock observations.

In Figure~\ref{fig:hauv}, we first show the \ha-to-UV flux ratios as functions of the observed phases for two SFH families that differ in fluctuation amplitudes.
For these cases, the \ha-to-UV flux ratios differ markedly. 
The larger amplitude leads to an increased scatter in \ha/UV, as expected. 
However, this difference becomes significantly less noticeable for models with the same fluctuation amplitude, but different durations and/or slopes.

The significant changes in the sSFRs for the more steeply rising SFH models drive the sudden change in \ha/UV near $\phi=\pi$. However, some scatter is present in this transition, making it appear smoother than one might expect from the abrupt changes in the simulated SFHs at $\phi=\pi$. The varying metallicities partly contribute to the smoothness. More importantly, the abrupt changes in the simulated SFHs are on the instantaneous sSFRs. As shown in \citet{Choi2017}, the number of ionization photons---and therefore---the \ha-to-UV flux ratios, is proportional to SFR averaged over 10 Myr, as opposed to instantaneous SFRs.

The simulated \ha/UV ratios are broadly consistent with the observationally inferred values reported in \citet{Asada2024}, although the maximum value of $\sim 0.07$ lies at the extreme end of their distribution. This upper limit also exceeds the maximum value found in \citet{Endsley2024}, likely due to differences in modeling assumptions, as \citet{Endsley2024} derived their \ha/UV ratios from \texttt{BEAGLE} \citep{Chevallard2016} fits to the photometry. We stress that our primary goal is not to precisely match the observations but rather to encompass a reasonable range of burstiness, since these simulations are designed to simply serve as a controlled baseline for evaluating the constraining power of different methods.

To quantify the constraining power of \ha-to-UV flux ratios, we compute the Wasserstein distance between the observed, noisy distributions to the model distributions. Also known as the earth mover's distance, the Wasserstein distance is a measure of the similarity between two probability distributions. It is similar to the perhaps more widely known Kullback-Leibler divergence, but is more robust to outliers and is sensitive to the spatial arrangement of the probability mass \citep{Panaretos2019}, and has been applied to astronomical data (e.g., \citealt{Li2024}).
It is possible to explore alternative distance metrics in the future, but here the best-fit model is found by simply identifying the shortest Wasserstein distance. 

We note that while the observed, noisy distributions are from the same model grid used to test the individual SED fits, the model distributions being compared to are generated from a denser grid ($\sigma \in $ [0.2, 0.9] dex with 0.1 dex increments, $\delta t \in $ [15, 45] Myr with 5 Myr increments, $\alpha \in $ [-0.3, -0.7] dex with 0.05 increments).
We repeat the mock observations with different realized noises and the resulting inference procedure 1,000 times for each SFH family sampled on the denser model grid as a robustness test.
We then calculate the fraction of population-level SFH parameters correctly identified as the best-fit, and consider the modal best-fit parameters as the overall best-fit.
For example, suppose we have a noisy distribution of \ha/UV corresponding to the SFH family of $\sigma=0.3, \delta t=40, \alpha=-0.4$. In the 1000 trials, 80\% of the time the shortest Wasserstein distance gives $\sigma=0.4$, 10\% of the time gives $\sigma=0.3$, and 10\% of the time gives $\sigma=0.5$. We take $\sigma=0.4$ as the best-fit, and plot it as a hexagon in Figure~\ref{fig:comp}. The shade indicates the possible range (i.e., from 0.3 to 0.5 in this example).

As seen from Figure~\ref{fig:comp}, the \ha-to-UV flux ratio almost always gets the large fluctuating amplitude right, and gets the small fluctuating amplitude quite close ($\sigma \sim 0.5$ dex when the truth is 0.4 dex). Notably, it never mistakenly infer one model with true $\sigma= 0.4$ dex as $\sigma \sim 0.8$ dex, and vice versa. This performance is markedly better than the individual SED fits, where the spread in the posterior medians can be larger than the difference between these two amplitudes. However, such discerning power vanishes when considering the other two parameters, namely duration and slope, as the best-fits only differ marginally for the different cases. Put another way, the \ha-to-UV flux ratio only has constraining power on the fluctuating amplitude: it is not very constraining for either the timescale of bursts or the underlying SFH slope.

These results demonstrate that the \ha-to-UV flux ratio is sensitive to recent bursts, which in turn can be linked to many physical processes driving the SFH, and that it is difficult to disentangle them based on the distribution of \ha-to-UV flux ratio alone. This challenge has also been implied in \citet{Mehta2023}, where they found that a smoothly rising or declining SFR over the long term can produce similar scatter in \ha-to-UV flux ratios as short-term bursts.

In summary, the \ha-to-UV flux ratio is a useful indicator for burstiness, but going beyond inferring a general characteristic of SFH variability (e.g., the fluctuation amplitude in our case) to constrain all the population-level SFH parameters requires additional information. This motivates us to explore a simultaneous population-level approach, as detailed below.

\section{A Simultaneous Population-level Approach to Constrain SFH Variations\label{sec:pop}}

The physical processes of various characteristic timescales influence certain spectral features in different ways. Figure~\ref{fig:pop} illustrates this with four examples of model SFHs and their corresponding spectra, and the spectral features of Balmer breaks, Balmer emission lines, near- and far-ultraviolet flux densities (NUV, FUV) are highlighted.
While isolated analysis of individual spectra is insufficient to robustly infer the population-level SFH parameters, simultaneous analysis of the distribution of the relevant spectral features may give greater discriminating power. 
This is because while determining the behavior of a single galaxy is challenging, but many galaxies living within the same space can effectively trace out phase-space to provide a comprehensive view of the population-level characteristics.

The above serves as a motivation for our proposed population-level approach; at the same time, it is also a key assumption that the formation histories of most galaxies vary on similar timescales. Here we show that galaxies drawn from the same SFH family can be used to identify that family. In real observations, one must somehow also first group galaxies into different SFH families. 

Simulations suggest that this may be challenging, as the probability of a burst (at fixed stellar mass) depends on a stability parameter, which involves gas mass and its specific angular momentum \citep{Cenci2024}. In other words, even at fixed mass the burstiness can depend on other galaxy properties such as their structure (see also \citealt{Hopkins2023}). However, as a first-order analysis, we may assume that galaxies in similar redshift and mass ranges likely experience comparable environmental conditions, leading to similar SFH variations.

\subsection{Spectral Features Sensitive to SFH Variations\label{subsec:features}}

\begin{figure*} 
 \gridline{\fig{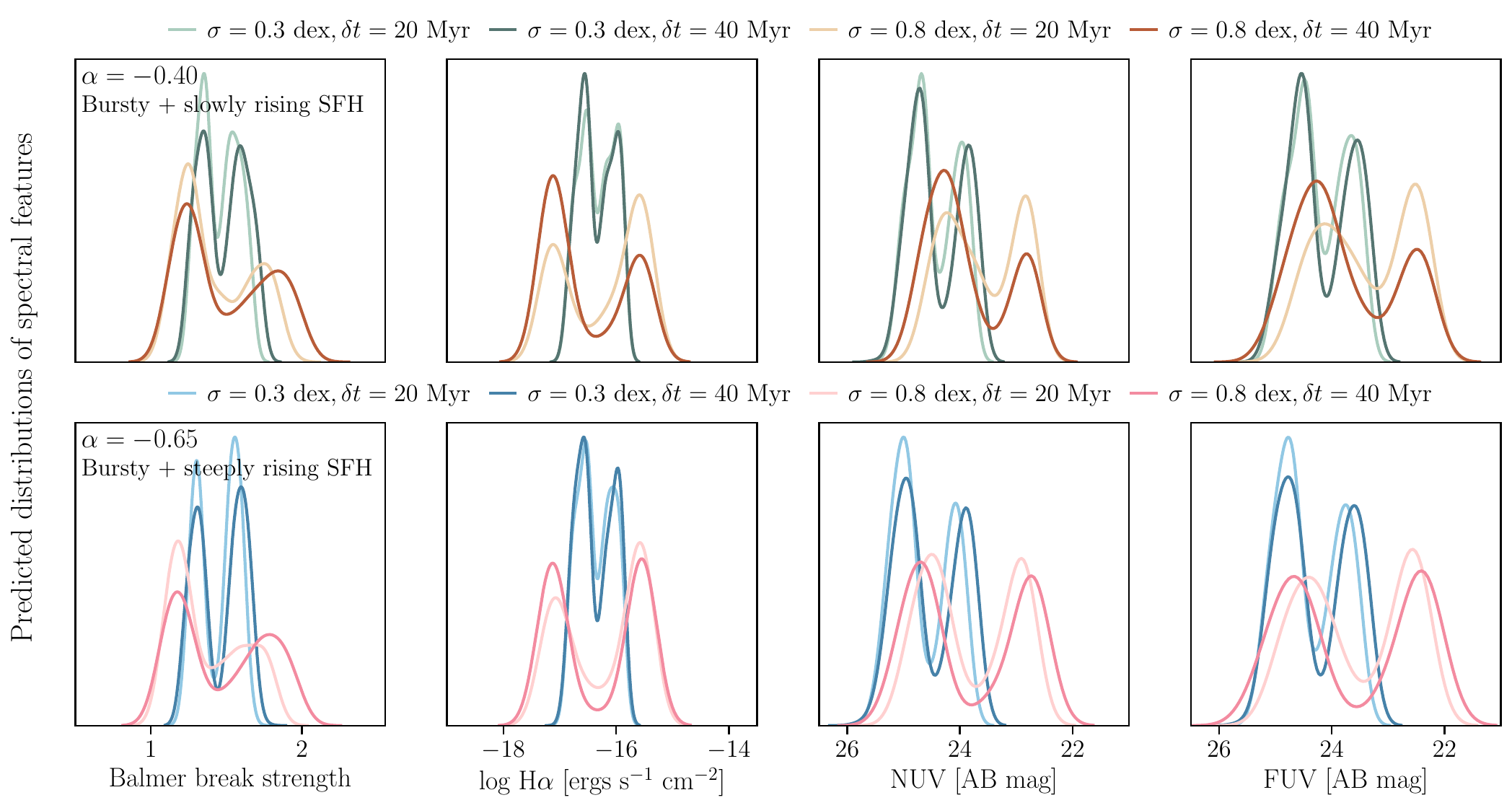}{0.95\textwidth}{}}
\caption{Proposed timescale-sensitive observables of this paper. The sampled population-level SFH parameters are the same as those used in testing the individual fits. Each row includes the predicted distributions drawn from SFH families with the same slope.
The distributions of the various spectral features are sensitive to different aspects of the SFH variability.}
\label{fig:pop_kde}
\end{figure*}

The suite of spectral features is chosen for their sensitivity to SFH variability over different timescales as described below, loosely following \citet{Iyer2024}.

\begin{enumerate}
  \item Balmer break strength: caused by the bound-free absorption by electrons in the electron band $n=2$, A-type ($1.4-2.1~\msun$) and later stars contribute to the break strength. We measure this from spectroscopy using the revised index from \citet{Wang2024:bb}, which uses two windows at [3620, 3720]~\AA\ and [4000, 4100]~\AA. This new definition captures the full Balmer break, as opposed to the 4000\,\AA\ break \citep{Balogh1999}. Critically, the 4000\,\AA\ break only appears after Gyr of quiescence, while the Balmer break appears after $\sim 100$~Myr. Given that the age of the universe at $z>4$ is $<1.6$~Gyr, a 4000\,\AA\ break is rarely seen. Our revised index is thus a more suitable age indicator at high redshifts.
  \item Balmer line emission: emerging from H~\textsc{ii} regions, this emission is indicative of the ionizing radiation from short-lived stars such as O and B-type stars ($5-10$~Myr). The readily available \ha\ emission is thus often taken as an ``instantaneous" SFR indicator.
  \item Dust-corrected NUV and FUV flux densities: due to outshining of the older stellar light by recent star formation, the UV in star-forming galaxies typically traces $30-70$~Myr timescales.
\end{enumerate}

We measure these spectral features from the simulated spectra (\S\,\ref{subsec:mock}). 
The resulting distributions for each of the population-level SFH parameters are shown in Figure~\ref{fig:pop_kde}.
We find that all the features are sensitive to the fluctuation amplitude, where a larger amplitude leads to a wider distribution. They are also moderately sensitive to the duration, which results in distributions peaking at different values.
Additionally, the dust-corrected NUV and FUV flux densities are sensitive to the slope, as either peaks become better separated, or their locations shift for a steeper slope.

It is worth emphasizing that most of the above distributions show wide spreads. This means that any one object, labeled with only one value from each observable, cannot trace the population distribution.

\subsection{Inferring Population-level SFH parameters\label{subsec:method_pop}}

Given the demonstrated sensitivity, then, in principle, the population-level description of burstiness (as parameterized by the population-level SFH parameters) could be fit to a population-level distribution of observables. As a proof of concept, we test whether the population-level distributions of observables can correctly identify the SFH parameters, sampled on the same model grid that the \ha-to-UV method is tested on in Section~\ref{sec:ha_uv}; that is, $\sigma \in $ [0.2, 0.9] dex with 0.1 dex increments, $\delta t \in $ [15, 45] Myr with 5 Myr increments, and $\alpha \in $ [-0.3, -0.7] dex with 0.05 increments). This analysis is justified on the basis that the spread in the posterior medians from the individual fits exceed the spacing of this model grid.

We begin by forward-modeling the observations corresponding to various SFH families (different combinations of the population-level SFH parameters), assuming Gaussian noise for each feature with S/N=20, and randomly generate a sample of 300 galaxies. 
Notably, the S/N here is generally noisier than the assumed individual spectra because each of these features is composed of multiple pixels, which are then averaged or combined to produce a spectral feature. This is especially true for FUV and NUV flux densities, which are the sum of many such pixels. We note that this setup represents a data quality level that is achievable in a JWST spectroscopic survey, e.g., UNCOVER \citep{Bezanson2022} and JADES \citep{Eisenstein2023}.

We then identify the best-fit model via the shortest Wasserstein distance in the same manner as outlined in Section~\ref{sec:ha_uv}, with a minor modification to account for the fact that four observables are considered here.
We treat each observable separately at first by calculating the Wasserstein distances of each, and then add the four Wasserstein distances in quadrature. The best-fit is again determined by this shortest total distance.

It is worth pointing out that we only consider the marginals here. Cross-correlations can in principle provide additional information. As the current setup already always identifies the correct population-level SFH parameters, we defer adding the cross-correlations to a future paper, where a full pipeline for the population-level fit will be presented.

\subsection{Assessing the Accuracy of Population Fits\label{subsec:res_pop}}

While the distribution of one or two observables alone may not always identify the correct population-level SFH parameter (e.g., the \ha/UV flux ratio discussed in Section~\ref{sec:ha_uv}), we find that by utilizing the four observables proposed in this paper, the correct SFH model can always be identified.

This performance is substantially more accurate and precise than the individual fits. Figure~\ref{fig:comp} shows the performance comparisons across all the methods considered in this paper in an illustrative manner. Specifically, the individual fits cannot distinguish between the sampled SFH parameters (e.g., 0.3 vs 0.8 dex fluctuation amplitudes) since the range in posterior medians from the individual fits exceeds the spacing of the sparse model grid. This justifies the simplified test on the population-level method in this work. Certainly, real galaxy populations reside in a more complex parameter space, and the next steps involve exploring the full density field, which will be detailed in a future paper.

\section{Discussion\label{sec:discussion}}

Having discussed the individual and population-level fits separately in Sections~\ref{subsec:ind_res}, \ref{sec:sed_fit_matchsfh}, \ref{sec:ha_uv}, and \ref{subsec:res_pop}, we now tie all the pieces together.
We begin this section by a brief summary, presenting the implications from this work for solving the longstanding challenge of outshining. Key lingering questions and future improvements are discussed in Section~\ref{subsec:diss:ques}.

The remaining two subsections supply additional discussion on the effects of priors placed on the SFH in individual SED fitting, and the implications of burstiness for studies assessing completeness.

\subsection{A Solution to Outshining?\label{subsec:diss:outshining}}

Inferring the past SFH from integrated light, where recent bursts dominate the SED---a phenomenon known as outshining \citep{Papovich2001}---remains a longstanding challenge. Outshining makes it inherently difficult to recover SFH behavior $> 100 - 200$~Myr in the past. The problem is further exacerbated in the high-redshift universe, where bursty SFHs are expected to be more common, complicating the inference of the behavior of SFR(t) even at timescales $< 100 - 200$~Myr.

Our analysis demonstrates that state-of-the-art inference methods with flexible SFHs and neutral priors systematically underestimate mass in bursty systems with rising SFHs by $\sim 0.15$ dex. While these methods reasonably recover average SFRs with a median offset of $\sim 0.1$ dex, they fail to capture the tens of Myr fluctuations in bursty SFHs. However, fitting a bursty SFH with the correct model for burstiness removes the biases: median offsets in mass and averaged SFR decrease to $\sim 0.04$ dex and $\sim 0.05$ dex, respectively (\S\,\ref{sec:sed_fit_matchsfh}). Furthermore, this approach facilitates the correct recovery of the detailed recent SFH.

The profound implication here is that the key to addressing outshining is to empirically measure the population distribution of bursty SFHs for a representative sample, and then apply this as an informed prior for SFR(t) for individual systems.
Note that here we attempt to solve outshining at a statistical level; that is, we do not claim to solve it for individual galaxies.

To this end, we introduce a population-level approach in Section~\ref{sec:pop}, aiming to recover the typical amplitude, timescale, and slope of the recent SFH to very high accuracy. Given the distributions of spectral features, we are able to correctly identify the corresponding SFH models via the shortest Wasserstein distance between the observed and the predicted distributions. By showing that an accurate model for SFH fluctuations does not suffer from the outshining problem, and that a model for SFH fluctuations can be accurately inferred from a complete and representative sample of galaxies, this work thus provides the first step towards solving outshining.

\subsection{Outstanding Questions\label{subsec:diss:ques}}

We elaborate on the lingering questions, along with possible ways forward, in this subsection.
First, in this work, we only sample a few representative points in the SFH model space, serving as a proof of concept. This choice is justified by the fact that the scatter in the part of the inferred SFH parameters from the individual approach already deviate from the truths by values larger than the spacing of the model grid, and that the scatters often exceed the spacing as well.
Certainly this is a simple test, but sufficiently illustrates that the population-level approach is a promising alternative. Going forward requires running a full fit exploring the complete parameter space, possibly adapting a more flexible and realistic SFH model, and obtaining the probability distributions of the inferred parameters. A toy model along these lines is presented in \citealt{Iyer2024}. Although this test is limited, as it was conducted on ensemble observations of only about 30 galaxies, it demonstrates that the constraints are sufficient to differentiate between the timescale parameters.
Another option is to perform Bayesian hierarchical modeling, as alluded to in \citet{Wan2024}. Fundamentally, we can consider our problem as attempting to learn about a population from many individual measurements, one that hierarchical models are designed to address.

Second, a critical assumption in using population-level burstiness model to solve outshining is that the population-level burstiness predicts past SFH. This holds true when the conditions for star formation evolve slowly, i.e., SFR(t) in the past few hundred Myr is similar to SFR(t) in the past few Gyr. This can ``solve'' outshining in a statistical sense, but not necessarily for individual galaxies; for example, individual galaxy bursts driven by external events (e.g., mergers or interactions) won't necessarily be captured by an improved prior, though they will be more properly marginalized over. It has been shown that merger-induced SFR is particularly important in more massive galaxies ($\mstar > 10^{10}~\msun$) at lower redshifts ($z = 0 - 1$) \citep{Cenci2024} using the FIREbox cosmological volume simulation \citep{Feldmann2023}.
However, such interactions are not the main driver of burstiness at lower stellar masses (see also \citealt{Hopkins2023:disc}); instead self-regulated star formation is more common here, which is well-suited to a population model. Therefore, at least to first-order, a measurement of the population distribution of bursty SFHs promises to mitigate the substantial uncertainty in interpreting the light of the galaxies where recent bursts dominate the SEDs.

Third, our parameterization of the population-level SFHs is simplified. This is meant to be illustrative. Rigorous tests are needed to confirm whether more complex forms of SFHs, e.g., fluctuations over many timescales, can still be accurately inferred with the proposed methodologies.
On a related note, we perform the population-level fits with simulations assuming no dust, as is done in \citet{Sun2024} where a semi-analytic framework is developed to model bursty SFHs.
While high-redshift, low-mass galaxies are expected to be less dusty in general, uncertainty in the dust attenuation when fitting for real galaxies may degrade the SFH recovery. Specifically, the best way forward is to have a Balmer decrement and UV flux in order to accuratly constrain dust attenuation---generally this means medium-resolution rest-optical spectroscopy alongside rest-UV photometry or spectroscopy.

Fourth, this work does not account for systematics in stellar mass and SFH recovery that arise from assumptions in SPS model choices. SPS models include multiple components, such as an IMF, stellar isochrones, and nebular physics (see \citealt{Conroy2013} for a review). Previous studies have demonstrated that inferred parameters can be sensitive to these fundamental modeling assumptions \citep{Pacifici2023,Whitler2023,Wang2024:sps}. The extent to which these assumptions influence our proposed population-level approach remains an open question.

Nevertheless, we emphasize that this paper presents a thorough examination of the constraining bursty SFHs via multiple approaches, and the findings suggest that a population-level burstiness model is a promising first step to solve outshining. We will explore the remaining questions in future work. Below, we supply additional discussions regarding the individual fits and observability and selection effects as implied by our burstiness model.

\subsection{Individual Inference for Galaxy Burstiness\label{subsec:ind}}

\subsubsection{The Challenges: Limited Information Content in Individual Spectra and a Complex Likelihood Surface\label{subsubsec:ind_chall}}

To start, while all the spectra are well fit, the complex variations in the input SFHs are rarely recovered based on individual spectra. This holds true even with high-quality data and no model mismatch. Here, no model mismatch means that the mock observations are created using the same stellar population synthesis as the forward model used in the SED fitting. Simply put, for a single galaxy, it is not possible to distinguish between the burst amplitude, duration, and an underlying smooth increase in the SFH.

The above finding suggests that the fundamental problem is the information content---the individual spectrum does not contain sufficient information to accurately infer the population-level SFH parameters. 
The additional test by including the full MIRI coverage suggests a similar problem of insufficient information content (the mean bias typically is improved by $0.1-0.2$~dex for the inferred masses, and $<0.06$~dex for the averaged SFRs; see Appendix~\ref{app:mass}).
High S/N medium resolution spectra can produce greater constraining power, but still insufficient to constrain the complex SFH behaviors, as demonstrated in Figure~\ref{fig:grating}.

Meanwhile, in light of recent developments in inferring SFHs from SED fitting with more flexible or more physically realistic models, it is reasonable to ask whether this individual approach could be improved by adopting these new formalisms. Examples include Dense Basis \citep{Iyer2019}, which constructs SFHs independent of any functional form using Gaussian processes, and stochastic priors \citep{Wan2024} based on power spectral density analysis \citep{Caplar2019}.
However, as we demonstrated in Section~\ref{sec:ind_fit}, a significantly different SFH from the input SFH can still fit the data equally well, even in our best-case scenario where no model mismatch is present. The fact that the input SFH can fall outside the posterior distribution of the inferred SFHs suggests that the likelihood space is full of local maxima which produce fits of similar overall quality---a challenge that more flexibility will exacerbate. Even with our simple models, the sampler struggles to locate the global maximum; otherwise, the true SFH should be included in the posterior distributions. This issue of missing the global maximum is further illustrated in Appendix~\ref{app:sampler} by the variations in inferred parameters found when fitting the same data multiple times. Making the SFH model more flexible would only add complexity to the likelihood surface, making it unlikely that SFH recovery would improve this way without a corresponding leap in sampling efficiency (e.g., through simulation-based inference; \citealt{Hahn2022,Khullar2022,Wang2023:sbi,Zhang2023,Ho2024}, or machine-learning-improved sampling (e.g., \citealt{Karamanis2022,Lange2023}).

\subsubsection{Continuity vs. Bursty Continuity Priors\label{subsec:priors}}

The above lists the key points regarding the individual fits, here we further elaborate on an interesting finding---namely, the different performances between the continuity and bursty continuity priors.
Previous works have shown that inferring SFH from fitting photometric data are influenced by prior choices \citep{Leja2019:sfh,Wang2023:sys}, and here we demonstrate that this can still be true when fitting high S/N spectra. 

Surprisingly, the continuity prior which favors smooth and constant SFHs better recovers masses (Appendix~\ref{app:mass}) as well as the average trends in SFHs.
As illustrated in Figure~\ref{fig:ind_sed}, the bursty continuity prior permits more dramatic changes in the SFH, and thus is more likely to show long-term deviations from the true average SFH. This is most evident in the case of the input SFH having a steeply rising slope, where the continuity prior captures the overall shape of the SFH, but the bursty continuity prior tends to put most star formation in a recent burst (i.e., making these objects form in a shorter period of time). Figure~\ref{fig:grating} illustrates this prior impact on the fitting of medium-resolution spectra, in which we see that the bursty continuity prior favors an immediate quench after a recent burst. 
These findings suggest that the bursty continuity prior is more prone to the classic outshining problem, where light from recent star formation eclipses that of older stellar populations \citep{Papovich2001}.

A puzzle here is that the bursty continuity prior, in principle, still favors constant SFH, as the expectation value of the ratio between the adjunct time bins is one. This means that the solution found by the continuity prior, which traces the average SFH better, should have higher prior probability. In addition, the likelihood for the data should also be higher for an SFH that is more similar to the true input SFH by design. However, the observed discrepancy between the results obtained with the two priors challenges this expectation.
To investigate, we compare the likelihood and probability values between fits using the two priors, and find small differences with no systematic trends. Moreover, model comparison based on Bayes factors \citep{Trotta2008} suggests that the data are agnostic regarding the choice of prior.
It is thus possible that by allowing more dramatic changes in the SFH, the bursty continuity prior leads to a more complex likelihood surface for the sampler to efficiently explore, potentially affecting the resulting inference.

It is worth emphasizing that the differences in performance depend on the forms of the SFH. While the combination of bursty and rising SFHs in this paper results in the worsened performance of the bursty continuity prior, it has been shown to be more effective at detecting rejuvenation events \citep{Park2024}. This is yet another example highlighting the importance of carefully considering the choice of prior based on the relevant physical scenarios.

With the addition of the full MIRI coverage, the outshining problem is partially mitigated. The $\sim 0.2$~dex systematic underestimation of the masses found when fitting the SEDs, assuming the continuity prior, is no longer present (see Appendix~\ref{app:mass}). The mean bias in the averaged SFR is decreased by $<0.06$~dex.
These improvements are likely due to the mass-to-light ratio exhibiting minimal variation in this wavelength range, particularly at rest 2~$\mu$m (e.g., \citealt{Kettlety2018}).
However, the MIRI bands are still insufficient to accurately trace the SFH variations from individual SED fits.

\subsection{Observability and Selection Effects\label{subsec:selection}}

\begin{figure*}
 \gridline{
 \fig{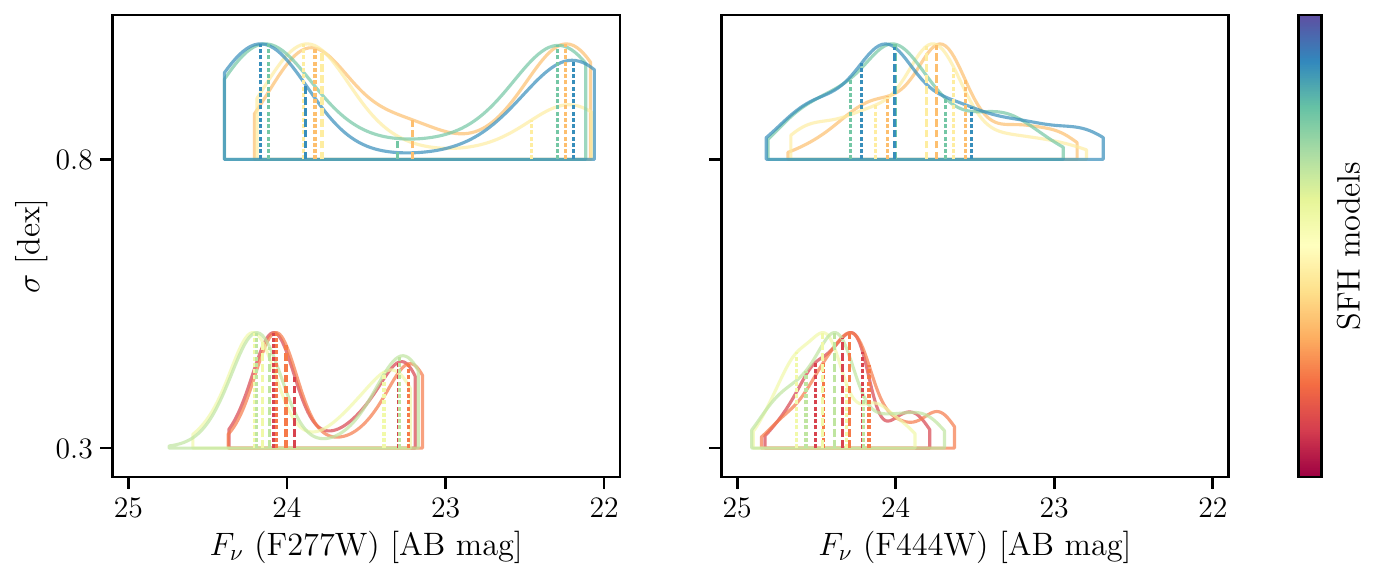}{0.95\textwidth}{}
 }
\caption{Population-level burstiness can shift the distributions of magnitudes at a fixed mass and redshift by $>2$ magnitudes. 
Fluctuation amplitudes in the SFHs are plotted as functions of observed fluxes in F277W band to the left, corresponding to rest-frame optical, and in F444W band to the right, corresponding to rest-frame near-infrared. Colors indicate the different families of SFH models, but the exact correspondence is irrelevant here. For a flux-limited survey, the observed sample of the galaxy population would most likely be dominated by those having larger fluctuation amplitudes (greater $\sigma$).
}
\label{fig:sel}
\end{figure*}

As hinted in Figure~\ref{fig:f444w}, SFH variations directly influence the observability of individual objects. For a galaxy population with the same mass and at the same redshift, different models for SFH variations alone produce over a magnitude of variability in observed fluxes in the F444W filter band, which corresponds to rest near-infrared. This is already the best-case scenario, since the fluxes in the rest near-infrared are expected to be the most stable. The variability in the F277W filter band, corresponding to rest-optical is even more dramatic, up to two magnitudes.

Figure~\ref{fig:sel} further connects the observability to various aspects of SFH variability by showing distributions flux densities for different population-level models of burstiness. For a flux-limited survey, the observed galaxy population near the observable limit would be dominated by those having SFHs with large fluctuation amplitudes (larger $\sigma$).
Although this variability in brightness is mainly driven by the bustier models with $\sigma=0.8$ dex, even with the more conservative case of $\sigma=0.3$ dex, the range of flux densities spans over a magnitude.

As pointed out in recent studies on observability using FIRE simulations \citep{Sun2023:seen} and a sample of Lyman-break galaxies \citep{Endsley2024}, our results also demonstrate that it is essential to quantify burstiness before completeness---the distribution of masses at a certain fixed flux level---can be assessed. This has further implications for determining stellar mass functions and global SFHs.
Changes on the SFMS due to mass incompleteness have already been shown in observations \citep{Leja2022} and in simulations \citep{McClymont2025}.

\section{Conclusions\label{sec:concl}}

This paper establishes the first step toward solving outshining---a longstanding challenge in interpreting the light from galaxies---by empirically measuring and then applying a model for short-term galaxy burstiness. This issue has become especially pertinent in light of recent observations from JWST of the early mini-quenched galaxies (e.g., \citealt{Looser2023:qg,Strait2023,Endsley2024,Weibel2024,Baker2025,Trussler2025}) and the over-abundance of luminous high-redshift galaxies (e.g., \citealt{Naidu2022,Atek2023,Casey2023,Finkelstein2023,Robertson2023}). 

We present a thorough examination of the extent of constraints attainable regarding bursty star formation, defined here as recurrent up-and-down fluctuations on tens of Myr timescales over the past 500 Myr of formation history.
We begin our study by assessing the state-of-the-art SED-fitting methods for inferring burstiness, and find that standard techniques with neutral priors fail to recover detailed recent SFHs and often under-estimate the total stellar mass; however, a successful recovery is achievable with the correct model for burstiness. We then introduce a population-level technique leveraging the distributions of timescale-sensitive spectral features, and demonstrate its potential for accurately inferring burstiness.
While motivated by high-redshift findings, our results are also applicable to studies of lower-redshift galaxies.
The main findings are summarized as follows.

First, in the presence of bursty star formation, the population-level SFH parameters, particularly the fluctuation amplitude, lead to strong selection effects in flux-limited surveys. The simulated brightness from our SFH models, for a galaxy population of the same redshift and mass, span over one magnitude in rest near-infrared, where the fluxes are expected to be the most stable. The variability in rest-optical is greater---more than two magnitudes. This may explain the prevalence of bursty star formation at high redshift in objects typically selected in the rest-optical, or in spectroscopic samples with more complex selection functions.

Second, even with exquisite data with no modeling systematics, individual SED fits are unable to place meaningful constraints on the population-level SFH parameters.
The spread in posterior medians in the inferred fluctuation amplitude, duration, and slope, from fitting an ensemble SEDs simulated from one SFH family, range from 0.5 to 1.0 dex, 10 to 120 Myr, and 0.5 to 1.2, respectively. These spreads are greater than the spacing of the model grid, meaning that the sampled SFH parameters cannot be distinguished. 
Critically, if these were real data, we would interpret the wide spread in posterior medians as evidence of a diversity in galaxy SFH timescales. This is especially concerning when the true SFH parameters fall outside the posterior coverage (Figures~\ref{fig:ind_sed}--\ref{fig:ind_sed_bursty}).

Third, the above findings are consistent for the tests done using the continuity and the bursty continuity priors. Surprisingly, the continuity prior performs marginally better than the bursty continuity prior. Assuming the continuity prior, the mean biases in the inferred masses, and SFRs averaged over the most 100~Myr are approximately $-0.06$~dex, and $0.1$~dex, respectively; assuming the bursty continuity prior, the biases increase to $-0.3$~dex and $0.2$~dex. This is because by preferring smooth transitions between time bins in the SFH, the continuity prior is less prone to outshining.

Fourth, the population distributions of spectral features---Balmer break strength, Balmer emission lines, NUV/FUV flux densities---are sensitive to at least some of the population-level SFH parameters. Considering these four observables simultaneously, we can always identify the correct population-level SFH parameters via the shortest Wasserstein distance. This work serves as a proof of concept, and the full methodology and validation of our proposed population-level method will be presented in a future work.

Finally, under the assumption that current population-level burstiness predicts past SFH, the complete formation history, even when the light from recent bursts dominates the SED, can be accurately inferred given the correct burstiness model.  In other words, the proposed population-level method for constraining burstiness is the key to address outshining. 

To conclude, bursty star formation has emerged as a potential unifying explanation for the high-redshift temporarily quenched galaxies and overly luminous galaxies discovered by JWST; yet observational constraints on burstiness remain uncertain.
Outshining represents perhaps the greatest challenge in interpreting the light from galaxies experiencing bursty star formation. The simultaneous population-level method proposed in this paper has the strong potential to constrain burstiness, and, thereby, addressing the issue of outshining. This work paves the way for a complete understanding of the observed diverse populations of galaxies.

\section*{Acknowledgments}
We thank Elia Cenci for providing the FIREbox SFHs.
We also thank Hiranya Peiris, Chris Hayward, and Kartheik Iyer for valuable discussions.
We are grateful to the anonymous referee for the helpful comments.
B.W. and J.L. acknowledge support from JWST-GO-02561.022-A.
Computations for this research were performed on the Pennsylvania State University's Institute for Computational and Data Sciences' Roar supercomputer. This publication made use of the NASA Astrophysical Data System for bibliographic information.

\software{
Dynesty \citep{Speagle2020}, Matplotlib \citep{2007CSE.....9...90H}, Nautilus \citep{Lange2023}, NumPy \citep{2020Natur.585..357H}, Prospector \citep{Johnson2021}, SciPy \citep{2020NatMe..17..261V}.
}

\section*{Appendix}
\counterwithin{figure}{section}
\counterwithin{table}{section}
\renewcommand{\thesection}{\Alph{section}}
\setcounter{section}{0}

\section{Effect of Re-binning SFHs in Sed Fitting\label{app:rebin}}

\begin{figure*}
 \gridline{
 \fig{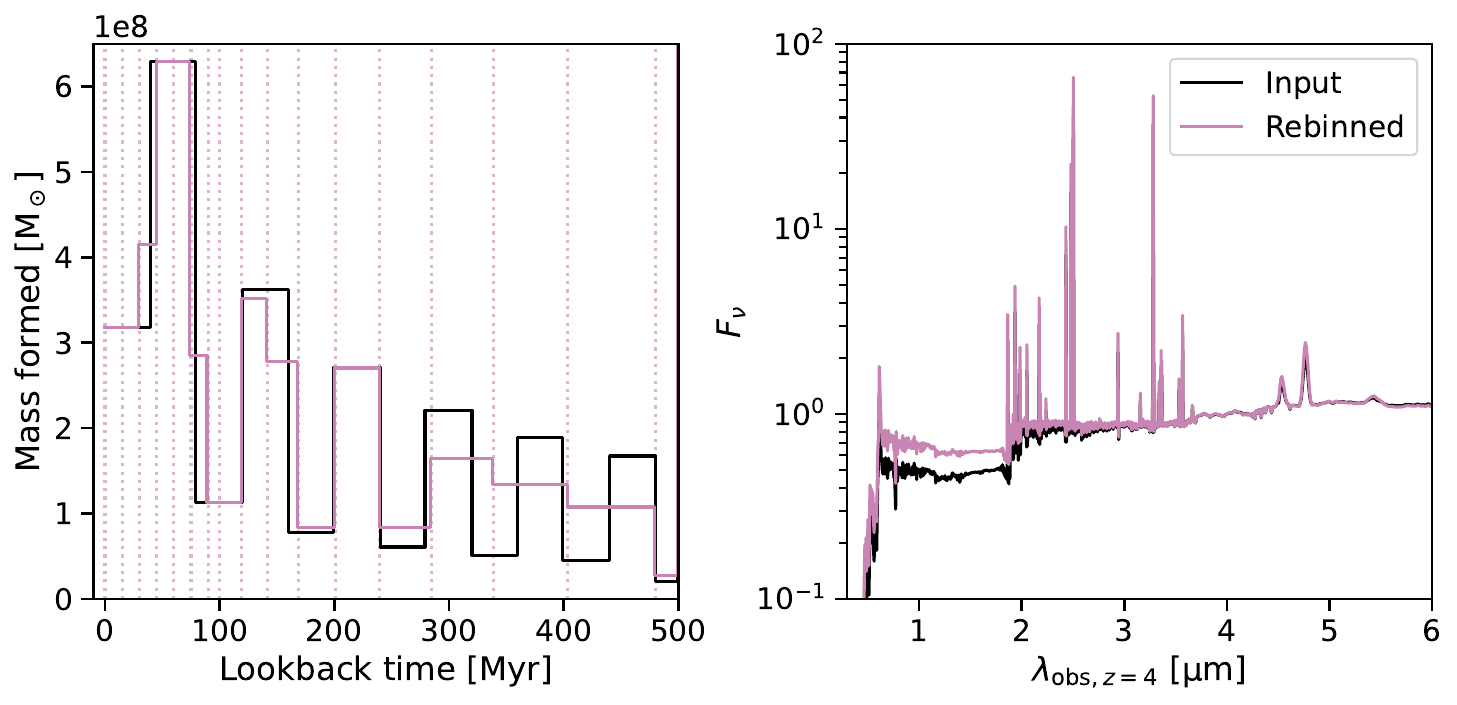}{0.95\textwidth}{}
 }
\caption{Rebinning of the model SFH changes the predicted spectrum. (Left) The SFH in intrinsic time bins is shown in black, whereas the rebinned SFH is shown in purple. (Right) Predicted spectra from the two SFHs are shown in the same colors.}
\label{fig:app:agebins}
\end{figure*}

In the main text, we adopt the same time bins as used in the input SFH when fitting SEDs. However, as it is difficult to know the timescales that the SFH varies a priori, certain form of rebinning is likely required.  

\citet{Leja2019:sfh} has tested different number of time bins extensively on smoothly varying SFHs, and concluded that rebinning has no impact on the SED-fitting results. We revisit this issue here given the more complex bursty SFHs tested in this work. We rebin the input SFH into 5 Myr fine bins in the first 100 Myr, and the rest into a uniform logarithmic grid with the last bin fixed to 90\% of the age of the universe. This choice is motivated by the need to be sufficiently flexible to describe SFHs potentially varying on a range of timescales, while keeping the total number of bins (and hence the total number of free parameters) to be within a manageable number. As shown in Figure~\ref{fig:app:agebins}, the input and the rebinned SFHs predict different model spectra. While alternative non-parametric SFH implementations exist (e.g., \citealt{Suess2022,Iyer2024}), we demonstrate, in this abbreviated manner, that the binning of SFH would need to be carefully selected when inferring SFH variability in SED fitting.

\section{Individual Fits across Different Data Resolutions using Flexible time bins\label{app:grating}}

\begin{figure*}
 \gridline{
 \fig{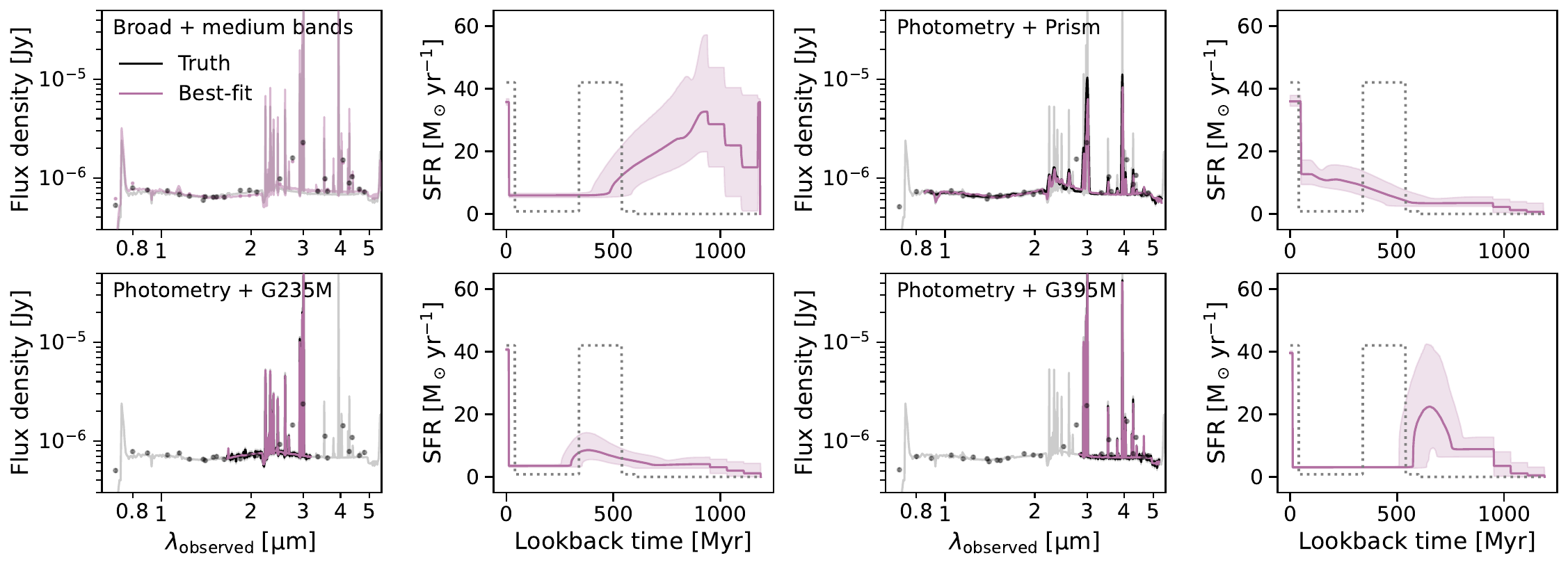}{0.95\textwidth}{}
 }
\caption{Inferring rejuvenation in the SFH from individual fits, using flexible time bins describing the SFH.
The plots are organized in the same format as Figure~\ref{fig:grating}. 
A short (200 Myr) period of past star formation cannot be accurately recovered, as the recent burst dominate the observed SED. Here we assume the ideal scenario where dust attention is known perfectly.
}
\label{fig:grating_rej}
\end{figure*}

We adopt the same general \prospector\ settings as outlined in Section~\ref{subsec:sed_fit}, but using flexible time bins \citep{Suess2022}, which allows the bin widths to change, so that the period where the SFH is inferred to have the most variations can be more described in greater resolution.
 
Four observing strategies are considered: photometry only (including broad and medium bands), photometry + low-resolution Prism spectroscopy, photometry + medium-resolution G235M spectroscopy, and photometry + medium-resolution G395M spectroscopy. At our redshift range of interest ($z \sim 4 - 7$), the G235M grating covers a critical age indicator---the Balmer break---and also H$\beta$, whereas the G395M grating covers H$\alpha$ and H$\beta$. 

The results, where we assume the ideal scenario of dust attention being known perfectly, are presented in Figure~\ref{fig:grating_rej}. 
This rejuvenating case---where short periods of star formation and quiescence occur in the recent past---produces a model spectrum in which the most recent burst dominate the observed SED. Specially, this galaxy experiences a period of star formation over the most recent 340--540 Myr, followed by a 200 Myr long quiescence, and then forms stars again during the most recent 40 Myr. 

Without the complication of the degeneracy between dust attention and emission-line strength, the most recent SFHs are inferred accurately in all cases. However, outshining induces significant uncertainty in inferring the past SFHs.

We also perform the same fits but letting dust attention to be fit for. The inferred becomes all consistent with a single burst, and the most recent SFRs are overestimated by a factor of $\sim 2$.

\section{Challenge in Mapping out the Likelihood Surface\label{app:sampler}}

\begin{figure} 
 \gridline{\fig{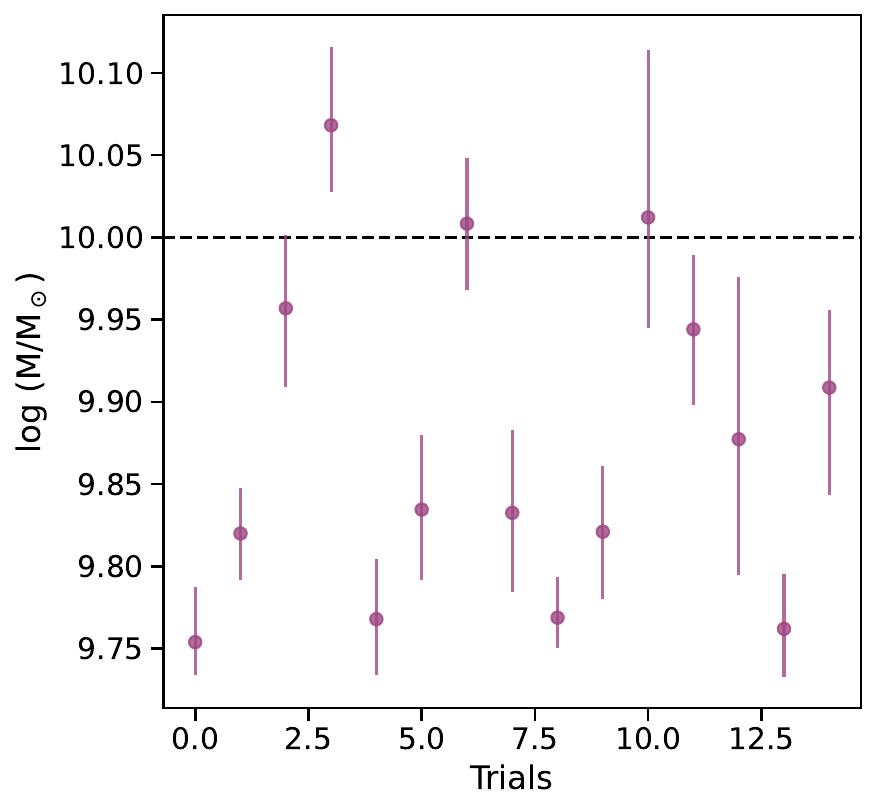}{0.45\textwidth}{}} 
\caption{Testing the sampler's ability to find the global maximum in the likelihood space. Total masses inferred from fitting the same SED multiple times are plotted in purple. The true mass is indicated by the black dashed line. The scatter as well as the changes in the uncertainties mean that locating the global maximum is not guaranteed.}
\label{fig:app:nruns}
\end{figure}

\begin{figure*}
\gridline{
 \fig{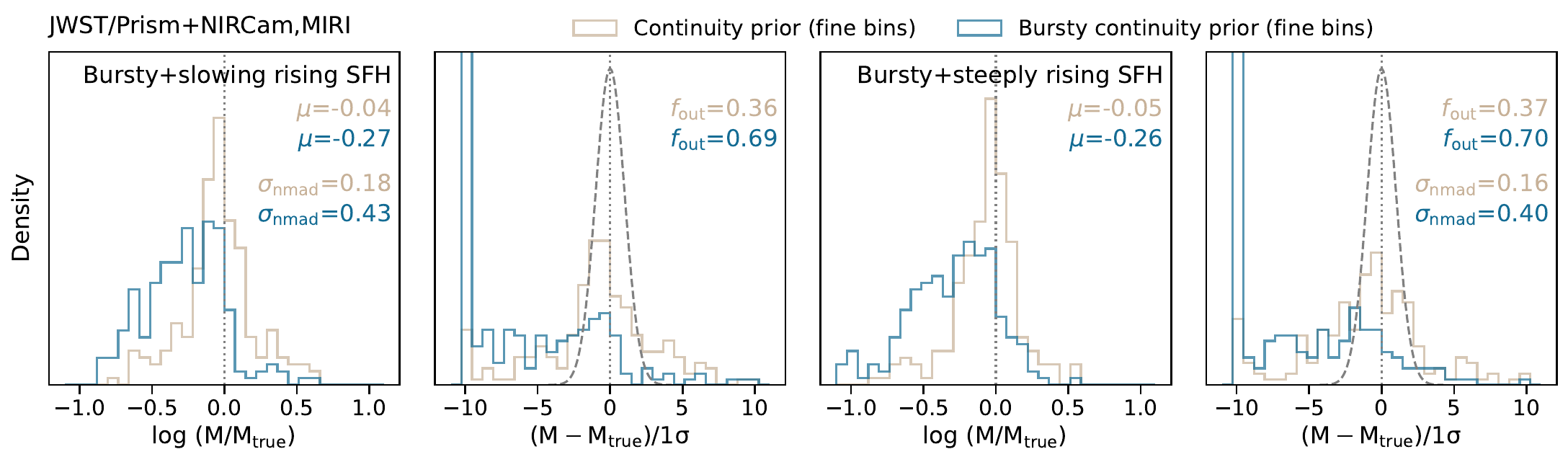}{0.95\textwidth}{(a)}
 }
\gridline{
 \fig{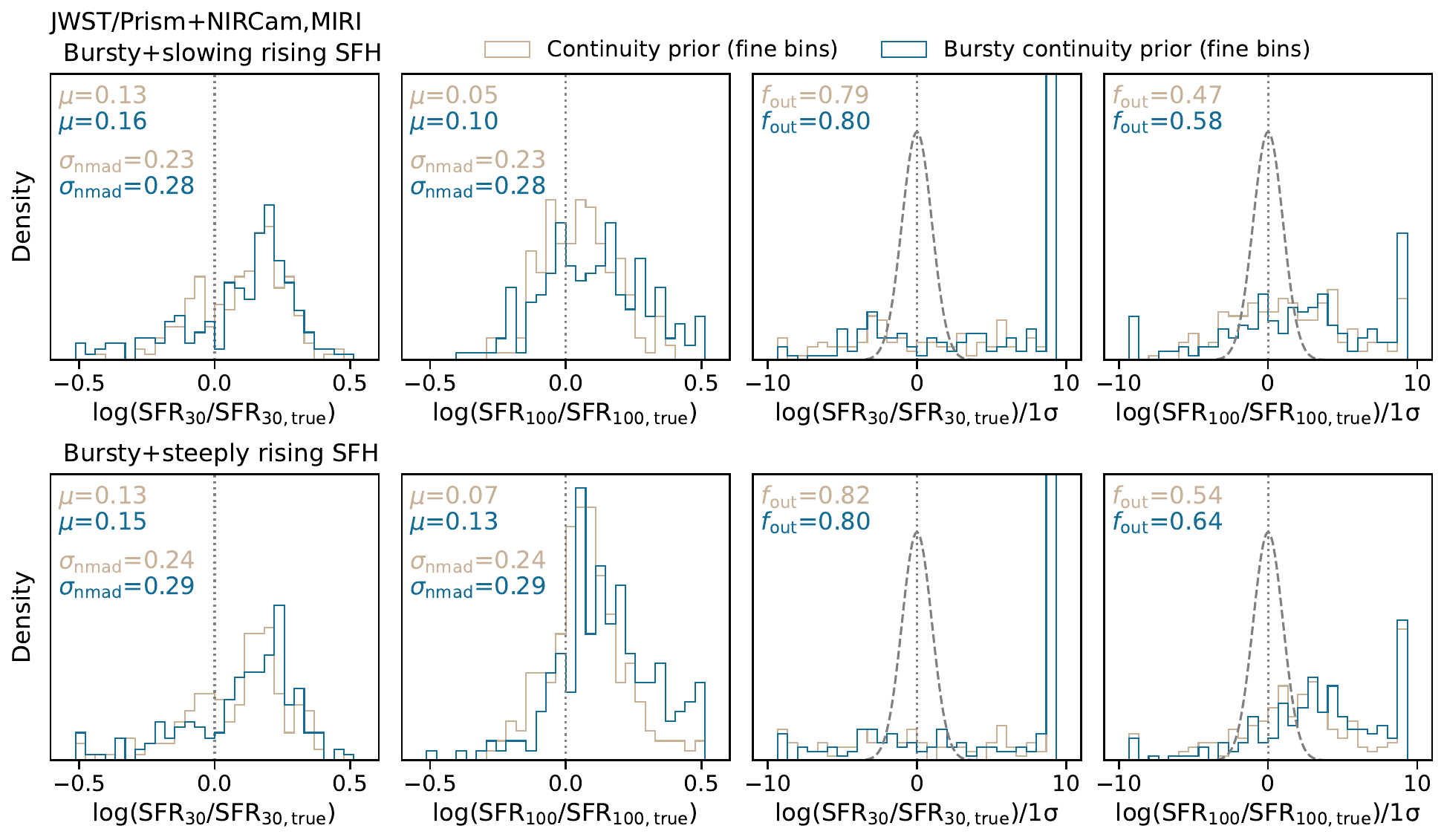}{0.95\textwidth}{(b)}
 }
\caption{The recovery of mass and SFR after including all the MIRI bands, shown in the same format as Figure~\ref{fig:sps_logm}. The biases in inferred mass see improvements at a $0.1-0.2$ dex level, whereas the improvements in the averaged SFRs are more marginal, $<0.06$ dex.
}
\label{fig:app:sps_logm}
\end{figure*}

The challenge being put forth by the complex likelihood surface is manifested in the underestimated uncertainties in Figures~\ref{fig:sps_logm} and \ref{fig:app:sps_logm}.
To verify the hypothesis that the sampler fails to locate the global maximum on the likelihood surface, we fit the same galaxy multiple times. The only randomness in this process is the noise---while the S/N is always set to 20, we let the realized noise to be generated randomly for each trial.
As seen from Figure~\ref{fig:app:nruns}, the inferred masses as well as their uncertainties can change significantly.
In addition, we repeat the same exercise using \texttt{nautilus}, which is also a nested sampler but with enhanced efficiency aided by neural networks \citep{Lange2023}. Interestingly, \texttt{nautilus} tends to consistently identify the underestimated mass as the final solution. 

A rigorous test of various samplers is out of the scope of this paper; however, our findings suggest that the sampling problem here is indeed a challenging one.
The bias in mass recovery is perhaps primarily driven by the challenge in finding the global maximum on the likelihood surface. High-dimensional SED modeling is known to have a complex likelihood space, and the complex behavior of bursty and rising SFHs exacerbates this challenge. As discussed in \citet{Wang2024:sps}, the nested sampling method \citep{Skilling2004}, which is also used in this paper, is already better suited to sample multi-modal posteriors than other traditional techniques such as Markov chain Monte Carlo \citep{Goodman2010}. In principle, the problem of locating the global maximal likelihood modes can be mitigated by substantially increasing in the accuracy settings, but the resulting increase in the CPU-hours would quickly become prohibitively expensive.

Possible improvements on the current state-of-art sampling process used in astronomy \citep{Feroz2009,Foreman-Mackey2013,Handley2015,Speagle2020,Buchner2021} include gradient-based samplers \citep{Hoffman2011}, combining with machine-learning techniques \citep{Karamanis2022,Lange2023}, or new inference workflows \citep{Hahn2022,Khullar2022,Wang2023:sbi,Zhang2023,Ho2024}.
It is worth noting that this missing-mode issue does not undermine the main conclusion in the paper that the individual spectrum does not contain enough information for complex recent SFH recovery. We have shown that all the spectra are well fit, but the inferred SFH can be very different from the input SFH. A sampler that is able to better map likelihood space can improve the uncertainty calibration, but unlikely to solve the fundamental problem regarding the information content.

\section{Effect of Adding JWST/MIRI in the Mass and SFR Recovery from SED Fitting\label{app:mass}}

The long-wavelength coverage enabled by MIRI improves the recovery of mass and averaged SFR. The bias typically is improved by $0.1-0.2$~dex for the inferred masses, and $<0.06$~dex for the SFRs. This is likely because the mass-to-light ratio shows minimal variation in this wavelength range, particularly at rest 2 $\mu$m (e.g., \citealt{Kettlety2018}).
However, the MIRI bands are still insufficient to accurately trace the SFH variations from individual SED fits.

\bibliography{sfh_wang.bib}
\bibliographystyle{aasjournal}

\end{CJK*}
\end{document}